\documentclass[12pt]{article}
\usepackage{amssymb}
\usepackage{graphicx}
\usepackage{float}
\usepackage{slashed}
\usepackage{amsmath}
\usepackage{tabularx}
\usepackage{breqn}
\usepackage{authblk}
\usepackage{array}
\usepackage{dsfont}
\usepackage{cite}
\usepackage{physics}
\usepackage[section]{placeins}
\usepackage[a4paper, total={6.8in, 10in}]{geometry}
\newcommand{\be}{\begin{equation}}
\newcommand{\ee}{\end{equation}}
\newcommand{\bea}{\begin{eqnarray}}
\newcommand{\eea}{\end{eqnarray}}
\newcommand{\ba}{\begin{aligned}}
\newcommand{\ea}{\end{aligned}}
\begin{document}
\title{Effect of gravity on Neutrino Oscillations in $\kappa$-deformed space-time}

\author{Harsha Sreekumar \thanks{harshasreekumark@gmail.com}, E. Harikumar \thanks{eharikumar@uohyd.ac.in}, }
\affil{School of Physics, University of Hyderabad, \\Central University P.O, Hyderabad-500046, Telangana, India}

\maketitle

\begin{abstract}

In this study, we analyse how quantisation of space-time affects propagating fermions in the presence of gravity. Effect of gravity is incorporated using spin connection which consists of universal torsion-free Levi-Civita connection and the Contortion tensor. This leads to the appearance of a four-fermion interaction term in the Lagrangian and because of non-commutativity of space-time, the deformation of the interaction term is found to depend on the background metric through tetrads. Further, we incorporate this interaction term to see the effect of gravitational interactions of neutrinos with the background matter in non-commutative space-time and study its effect on neutrino oscillation probabilities.
\end{abstract}

\section{Introduction}

Neutrinos were first postulated by Pauli in 1930 to account for the continuous energy spectrum associated with $\beta$-decay and the theory was experimentally verified by Cowan and Reines in 1953\cite{clyde,frederick}. Later, Standard Model of particle physics, a unified description of three of the four fundamental forces was developed. Although the Standard Model is well-celebrated and most of its predictions are experimentally verified, one of the drawbacks of the model is that it cannot account for the observed mass of the neutrinos. Fermions acquire mass after spontaneous symmetry breaking via the Higgs mechanism where the mass term is: $-y\overline{\psi}_{L}\psi_{R}\left<\phi\right>$. Here, $y$ is the Yukawa coupling, $\psi_{L}, \psi_{R}$ are the left-handed and right-handed fermion fields, respectively and $\left<\phi\right>$ is the vacuum expectation value of the Higgs field. Since there are no right-handed neutrinos in the Standard Model, neutrinos are massless within the Standard Model. But Neutrino Oscillation is an experimentally observed quantum mechanical phenomenon where a specific flavour of neutrino changes to neutrinos of a different flavour. Neutrinos have to be massive for the occurrence of this phenomenon\cite{pontecorvo} and they are always produced in flavour eigenstates which is a coherent superposition of the mass eigenstates, $\nu_{i}$. Thus, a neutrino of definite flavour is a linear combination of states of definite mass and as they propagate, the phase between the mass eigenstates changes with distance. This leads to a difference in relative phase between the mass eigenstates as the neutrinos propagate and seen at the detector which could result in the detection of a different flavour state compared to the flavour state created at the source. Main sources of neutrinos are the sun, supernova explosions, interactions of cosmic rays with the atmospheric nuclei and from nuclear reactors and accelerators. The phenomenon of neutrino oscillation has been verified by experiments involving solar, atmospheric, accelerator and reactor neutrinos. The Solar neutrino problem where the observed neutrino flux from the Sun is inconsistent with the theoretical prediction was resolved by the observation of neutrino oscillations through SNO experiment \cite{SNO1,SNO2} and Kamiokande experiment \cite{kamiokande}.

Neutrino Oscillations have been extensively studied in the presence of gravity. In \cite{ahluwalia}, the quantum mechanical neutrino-oscillation phases that is induced gravitationally is discussed and their observation possibilities for type-$\uppercase\expandafter{\romannumeral 2}$ supernova are considered. Here, the evolution of the neutrino from the source region to the detector region is written using the time translation operator, the Hamiltonian, and the space translation operator or the momentum operator. In the weak gravitational field limit, the time evolution of these operators are rewritten in terms of $h_{\mu\nu}=g_{\mu\nu}^{W}-\eta_{\mu\nu}$, where $g_{\mu\nu}^{W}$ is the Schwarzschild space-time metric in the weak field limit and $\eta_{\mu\nu}$ is the Minkowski metric. This leads to a new gravitationally induced quantum mechanical phase arising in the oscillation probability equations along with the usual kinematic phase. The gravitationally induced phase and the kinematic phase are related to each other through the average gravitational potential. This effect is expected to be observed in astrophysical environments and hence the gravitationally induced phase and its effect on the oscillation probabilities are calculated for supernova explosions. Neutrino Oscillations in curved space-time is studied in \cite{cardell} where gravitational effects on the oscillations were found to be related to the redshift. The time evolution operator is generalised to curved space-time along with the inclusion of matter effects. The neutrino spin flips because of the gravitational fields and this effect is incorporated using the spin connection which is written in terms of gamma matrices and tetrads. These analyses are further studied for Schwarzschild geometry where it was observed that contributions of gravity to spin precession vanish. However, these effects were found to be closely related to gravitational redshift. In \cite{forengo}, propagation of neutrinos in gravitational field was studied and the obtained results were further extended to Schwarzschild metric. An additional quantum-mechanical phase was calculated for neutrinos in curved space-time and this new phase is analysed for neutrinos propagating in the radial direction as well as in the non-radial direction in Schwarzschild metric. Further, gravitational lensing of the neutrinos is studied and neutrino transition probabilities are derived in \cite{forengo}. In \cite{amlicar,amlicar2}, propagation of neutrinos is studied in the exterior part of a black hole in the Kalb-Ramond background. Effect of spontaneous violation of Lorentz symmetry due to a black hole solution on the propagation of neutrinos is studied in the context of bumblebee gravity in \cite{amlicar3}.

Gravity is the weakest of the four fundamental forces and the least understood in the microscopic scale. Einstein's general theory of relativity is extremely successful in describing macroscopic effects of gravity but the theory breaks down in the microscopic realm. Although gravitational interactions of elementary particles is extremely weak and currently immeasurable with the present technology, they cannot be neglected when studying phenomena involving ultra high density matter such as provided by a collapsing star and in situations when the gravitational effects are strong. Thus extension of general theory of relativity to the microscopic realm is essential in understanding the gravitational interaction of elementary particles \cite{hehl}.

Mass and spin angular momentum play an essential role in describing the behaviour of elementary particles and their interactions. In macroscopic realm, mass is associated with the energy-momentum tensor while the total spin averages out due to its dipole nature. General theory of relativity connects matter to the Riemannian space-time geometry by considering mass or energy- momentum tensor to be the source of gravitational field. In the microscopic realm, spin angular momentum along with the mass has to be intrinsically connected to the space-time geometry. Thus, to introduce the effect of spin, one has to move from Riemann space-time to Riemann-Cartan space-time. Here, the non-Riemannian part of the affine connection is known as the contortion tensor which is the geometrical quantity with rotational degrees of freedom that couples the spin angular momentum to the space-time geometry\cite{hehl}.

An effective quartic interaction term is found to exist in the Lagrangian when contortion part of the spin connection is considered along with the universal gravitational Levi-Civita connection when studying the propagation of fermions in curved space-time \cite{amitabha}. This quartic interaction term has coupling constants which, if not too small, will affect the neutrinos passing through normal matter. The geometrical contribution due to gravitational effects on neutrinos passing through matter is calculated and neutrino oscillation probabilities are derived. This study is extended to include Preliminary Reference Earth Model(PREM) and to solve the Schr$\ddot{o}$dinger equation to study the effect of gravitation on $\nu_{\mu} \rightarrow \nu_{\tau}$, $\nu_{\mu} \rightarrow \nu_{e}$ transition probability and $\nu_{\mu} \rightarrow \nu_{\mu}$ survival probability \cite{amitabha2}. The effect of torsion is considered to be more noticeable for atmospheric neutrinos. The dependence of oscillation probability for atmospheric neutrinos on the CP phase angle is also studied in the presence of gravitation in \cite{amitabha2}. Neutrinos are very useful to study the internal structures of supernova and the high matter density inside a core collapse supernova is found to affect the four-fermion interaction appearing because of gravitational interactions of fermions \cite{amitabha3}. It has been shown that the $\nu_{e} \rightarrow \nu_{e}$ survival probability is periodic but the coupling constant affects the position of its first dip \cite{amitabha3}. The effect of the coupling constant appearing in the effective Hamiltonian due to the gravitational interaction of the fermions in DUNE and P2SO experiments is studied in \cite{papia}. It is found that P2SO gives a more rigid bound on the coupling constants than DUNE and when both the coupling constants are considered for three flavours of neutrinos, the upper bound on them weakens \cite{papia}.

Although Standard Model successfully unifies three of the four fundamental forces, it leaves out gravitational force, as a consistent quantum theory of gravity remains a mystery. There are numerous approaches to construct quantised description of gravity and a common feature in all these approaches is the existence of a fundamental length scale. Quantised space-time or non-commutative space-time is one of the approaches motivated by Heisenberg's Uncertainty principle and General Theory of Relativity where space-time coordinates become operators \cite{connes,dop}. The fundamental length scale inherent in all models of quantum gravity appears naturally through the commutation relation between the coordinates in non-commutative space-time, which makes it an interesting model to study quantum gravity effects.

In this paper, we study the effect of gravity on propagating fermions and how this gravitational interaction affects neutrino oscillation probabilities in non-commutative space-time. Section $2$ gives an overview of $\kappa$-deformed space-time which is a type of non-commutative space-time. The effect of gravity on propagating fermions in $\kappa$-deformed space-time is discussed in Section $3$, where we derive the effective Lagrangian after incorporating gravitational interaction through the spin connection. We also derive the corresponding Dirac equation of the fermions in this section and look into the four-fermion interaction term appearing in the Lagrangian because of the gravitational interaction of fermions in $\kappa$-deformed space-time. Section $4$ discusses two- flavour neutrino oscillation probabilities in vacuum in $\kappa$-deformed space-time as well as in matter after incorporating the effects of gravity through the four-fermion interaction in $\kappa$-deformed space-time. Three-flavour neutrino oscillation probability in the presence of gravity in $\kappa$-deformed space-time is discussed in Section $5$ and they are further analysed and plotted in Section $6$ in the Schwarzschild metric background. Section $7$ presents the concluding remarks. An appendix is added to discuss how tetrads and gamma matrices are generalised to non-commutative space-time.

\section{$\kappa$-deformed space-time.}

$\kappa$-deformed space-time is a Lie-algebraic type non-commutative space-time whose coordinates satisfy the commutation relation,
\be
[\hat{x}^i,\hat{x}^j]=0,~~~[\hat{x}^0, \hat{x}^i]=ia\hat{x}^i,~~~a=\frac{1}{\kappa}. \label{ksp-1}
\ee
Here, $a$ is deformation parameter with the dimension of length. Field theoretical models in $\kappa$-deformed space-time can be constructed using the star product formalism where the usual pointwise multiplication between the coordinates or fields is replaced by a star product \cite{das,dimitri}. $\kappa$-Poincare algebra is the associated symmetry algebra of this star product formalism. Another method to construct field theoretical models in $\kappa$-space-time is by defining the non-commutative coordinates in terms of commutative variables and their derivatives using different realizations \cite{meljanac,meljanac2}. This approach is equivalent to the star product formalism and different realizations of non-commutative variables correspond to different orderings in the star product formalism \cite{meljanac3}. We have used the realization approach in our study, namely, the $\varphi$ realization, where the non-commutative coordinates are written as,
\be \label{ksp-2}
 \hat{x}_0=x_0\psi(ap^{0})+iax_j\partial_j\gamma(ap^{0});~~~~\hat{x}_i=x_i\varphi(ap^{0}),
\ee
Here, $\psi(ap^{0}),~\gamma(ap^{0})$ and $\varphi(ap^{0})$ satisfy the conditions,
\be \label{ksp-3}
\psi(0)=1,~\varphi(0)=1,~ \gamma(0)=\varphi^{\prime}(0)+1.
\ee
Note that $p^{0}$ is the energy of the particle probing the space-time and in the further discussions, we refer to $ap^{0}$ as the deformation parameter. By Substituting eq.~\eqref{ksp-2} in eq.~\eqref{ksp-1}, one obtains,
\be \label{ksp-4}
\frac{\varphi^{\prime}(ap^{0})}{\varphi(ap^{0})}\psi(ap^{0})=\gamma(ap^{0})-1.
\ee
Here, $\varphi^{\prime}(ap^{0})=\frac{d\varphi}{d(ap^{0})}$. Of all the allowed choices of $\psi, \gamma$ and $\varphi$, we choose $\psi(ap^{0})=1, \gamma(ap^{0})=0$ and $\varphi(ap^{0})=e^{-ap^{0}}$ \cite{meljanac,meljanac2} which reduces eq.~\eqref{ksp-2} as, 
\be\label{ksp-5} 
\hat{x}_{0}=x_{0},~~~\hat{x}_{i}=e^{-ap^{0}}x_{i}.
\ee

The symmetry algebra of $\kappa$-Minkowski space-time in the realisation approach is the undeformed $\kappa$-Poincare algebra whose generators satisfy \cite{meljanac,meljanac2},
\be \label{ksp-generator} 
\left[M_{\mu\nu}, M_{\lambda\rho}\right]=M_{\mu\rho}\eta_{\nu\lambda}-M_{\nu\rho}\eta_{\mu\lambda}-M_{\mu\lambda}\eta_{\nu\rho}+M_{\nu\lambda}\eta_{\mu\rho}.
\ee
The explicit form of the generators of the undeformed $\kappa$-Poincare algebra is,
\bea \label{ksp-generator2}
M_{ij}&=&x_{i}\partial_{j}-x_{j}\partial_{i},\\ \nonumber
M_{i0}&=&x_{i}\partial_{0}\left(\varphi\frac{e^{2ap^{0}}}{2ap^{0}}\right)-x_{0}\partial_{i}\left(\frac{1}{\varphi}\right)+iax_{i}\partial_{k}^{2}\left(\frac{1}{2\varphi}\right)+iax_{k}\partial_{k}\partial_{i}\left(-\frac{\gamma}{\varphi}\right).
\eea
It was found that the partial derivative $\partial_{\mu}=\frac{\partial}{\partial x^{\mu}}$ doesn't transform as a $4$-vector under the undeformed $\kappa$-Poincare transformation and hence a new derivative called the Dirac derivative, $D_{\mu}$ was introduced \cite{meljanac}. Under the $\kappa$-Poincare transformation, the Dirac derivative transforms as a $4$-vector, i.e.,
\bea \label{ksp-dirac derivative}
\left[M_{\mu\nu}, D_{\lambda}\right]&=&D_{\mu}\eta_{\nu\lambda}-D_{\nu}\eta_{\mu\lambda}, \\\nonumber
\left[D_{\mu}, D_{\nu}\right]&=&0 .
\eea
The explicit form of the components of Dirac derivative are, 
\bea \label{ksp-dirac derivative3}
D_{0}&=&\partial_{0}\left(\frac{sinh(ap^{0})}{ap^{0}}\right)+ia\partial_{k}^{2}\left(\frac{e^{-ap^{0}}}{2\varphi^{2}}\right), \\\nonumber
D_{i}&=&\partial_{i}\left(\frac{e^{-ap^{0}}}{\varphi}\right).
\eea

As one of the approaches to study quantum gravity effects is to use non-commutative space-time, it is important to analyse the effect of non-commutative space-time on interactions of elementary particles. We take up this problem here and study how $\kappa$-deformation of the space-time affects neutrino oscillations, to the first order in the deformation parameter $ap^{0}$.
 
\section{Gravitational Interaction of fermions in $\kappa$-deformed space-time.} 

In this section, we study the effect of gravity on fermions propagating in $\kappa$-deformed space-time.
The action of the fermionic field in the presence of gravity in non-commutative space-time is given as;
\be \label{deformedaction}
\hat{S}=\int \abs{e}d^{4}x\left[ \frac{1}{2K}\hat{e}_{a}^{~\mu}\hat{e}_{b}^{~\nu}\hat{F}_{\mu\nu}^{~ab}+\hat{L}_{\psi}\right],
\ee
where  $K=8\pi G$, \footnote{In eq.~\eqref{deformedaction}, the term $|\hat{e}|d^{4}\hat{x}$ in the integral reduces to $|e|d^{4}x$ as $|\hat{e}|=e^{3ap^{0}}|e|$ and $d^{4}\hat{x}=e^{-3ap^{0}}d^{4}x$.} and
\be \label{fieldstrengthtensor}
\hat{F}_{\mu\nu}^{~ab}=D_{\mu}\hat{A}_{\nu}^{~ab}-D_{\nu}\hat{A}_{\mu}^{~ab}+\hat{A}_{\mu~~c}^{~~a}\hat{A}_{\nu}^{~cb}-\hat{A}_{\nu~~c}^{~~a}\hat{A}_{\mu}^{~cb}.
\ee
Here $D_{\mu}$ is the Dirac derivative in $\kappa$-deformed space-time which takes the form $D_{0}=\partial_{0}+\frac{ia}{2}\nabla^{2}$ and $D_{i}=\partial_{i}$ in the realisation we use\cite{meljanac}. Also, $\hat{A}_{\mu}^{~ab}=\hat{\omega}_{\mu}^{~ab}+\hat{\Lambda}_{\mu}^{~ab}$ is the deformed spin connection living in the $\kappa$-space-time where $(\hat{\omega}_{\mu}^{~ab})$ is the deformed torsion free Levi-Civita connection and $(\hat{\Lambda}_{\mu}^{~ab})$ the deformed contortion tensor. $\hat{\omega}_{\mu}^{~ab}$ gives the usual Riemannian metric $g$ and affine connection $\Gamma$ in deformed space-time while incorporation of $\hat{\Lambda}_{\mu}^{~ab}$ is necessary to study fermionic interactions in curved space-time to account for the spin of fermions. Note that the Latin indices $a, b,$ etc. denote quantities in the flat space-time, while the Greek indices $\mu, \nu$ etc. denote quantities in the curved space-time.

The Lagrangian of the fermionic field in the presence of gravity in $\kappa$-deformed space-time appearing in eq.~\eqref{deformedaction} is,
\bea \label{lagrangian1}
\hat{L}_{\psi}&=&\frac{i}{2}\left[\hat{\overline{\psi}}\hat{\gamma}^{a}\hat{D}_{a}\hat{\psi}-(\hat{\overline{\psi}}\hat{\gamma}^{a}\hat{D}_{a}\hat{\psi})^{\dagger}\right]-\hat{m}\hat{\overline{\psi}}\hat{\psi}, \\ \label{lagrangian2}
&=&\frac{i}{2}\left[\hat{\overline{\psi}}\hat{\gamma}^{\mu}D_{\mu}\hat{\psi}-D_{\mu}\hat{\overline{\psi}}\hat{\gamma}^{\mu}\hat{\psi}-\frac{i}{4}\hat{A}_{\mu}^{~ab}\hat{\overline{\psi}}[\hat{\sigma}_{ab}, \hat{\gamma}^{c}]_{+}\hat{\psi}\hat{e}_{c}^{~\mu}\right]-\hat{m}\hat{\overline{\psi}}\hat{\psi}.
\eea
where $\hat{D}_{\mu}=D_{\mu}-\frac{i}{4}\hat{A}_{\mu}^{~ab}\hat{\sigma}_{ab}$ is the Lorentz covariant derivative generalised to $\kappa$-deformed space-time, $\hat{\sigma}_{ab}=\frac{i}{2}[\hat{\gamma}_{a},\hat{\gamma}_{b}]_{-}$ and $\hat{\psi}=\hat{\psi}(\hat{x})$. In order to obtain eq.~\eqref{lagrangian2}, the explicit form of $\kappa$-Lorentz covariant derivative ($\hat{D}_{\mu}$) is substituted in eq.~\eqref{lagrangian1} and the identities $\hat{\gamma}^{0}(\hat{\gamma}^{a})^{\dagger}\hat{\gamma}^{0}=\hat{\gamma}^{a},~(\hat{\gamma}^{a})^{\dagger}\hat{\gamma}^{0}=\hat{\gamma}^{0}\hat{\gamma}^{a}$ and $(\hat{\gamma}^{b}\hat{\gamma}^{c})^{\dagger}\hat{\gamma}^{0}=-\hat{\gamma}^{0}(\hat{\gamma}^{b}\hat{\gamma}^{c})$ are used. Also, deformed tetrads are used to transform derivatives and gamma matrices of flat Minkowski space-time to $\kappa$-deformed curved space-time. In the flat space-time, the above Dirac Lagrangian reduces to the $\kappa$-Dirac theory discussed in \cite{EH}.

Substituting the deformed field strength tensor, $\hat{F}_{\mu\nu}^{~ab}$ in eq.~\eqref{fieldstrengthtensor} into eq.~\eqref{deformedaction}, we get,
\begin{multline}  \label{action2}
\hat{S}=\int \abs{e}d^{4}x\Big[\frac{1}{2K}\Big(\hat{R}(\hat{\Gamma})+D_{\mu}\hat{\Lambda}_{\nu}^{~ab}-D_{\nu}\hat{\Lambda}_{\mu}^{~ab}+\hat{\omega}_{\mu~~c}^{~~a}\hat{\Lambda}_{\nu}^{cb}-\hat{\omega}_{\nu~~c}^{~~a}\hat{\Lambda}_{\mu}^{cb}+\hat{\Lambda}_{\mu~~c}^{~~a}\hat{\omega}_{\nu}^{~cb}-\hat{\Lambda}_{\nu~~c}^{~~a}\hat{\omega}_{\mu}^{~cb}\\+\hat{\Lambda}_{\mu~~c}^{~~a}\hat{\Lambda}_{\nu}^{~cb}-\hat{\Lambda}_{\nu~~c}^{~~a}\hat{\Lambda}_{\mu}^{~cb}\Big)\hat{e}_{a}^{~\mu}\hat{e}_{b}^{~\nu}+\hat{L}_{\psi}\Big],
\end{multline}
where $\hat{R}(\hat{\Gamma})=D_{\mu}\hat{\omega}_{\nu}^{~ab}-D_{\nu}\hat{\omega}_{\mu}^{~ab}+\hat{\omega}_{\mu~~c}^{~~a}\hat{\omega}_{\nu}^{~cb}-\hat{\omega}_{\nu~~c}^{~~a}\hat{\omega}_{\mu}^{~cb}$ is the torsion free Levi-Civita part of the affine connection generalised to $\kappa$-deformed space-time. The metricity condition in $\kappa$-deformed space-time i.e., $\hat{\nabla}_{\mu}\hat{e}_{a}^{~\mu}=0$ takes the form,
\be \label{metricitycondition} 
D_{\mu}\hat{e}_{a}^{~\nu}=\hat{\omega}_{\mu~~a}^{~~c}\hat{e}_{c}^{~\nu}-\hat{\Gamma}_{\mu~~\sigma}^{~~\nu}\hat{e}_{a}^{~\sigma},
\ee
which is used to reduce the above action to 
\be \label{action3}
\hat{S}=\int \abs{e}d^{4}x\left[\frac{1}{2K}\left(\hat{R}(\hat{\Gamma})+\hat{\Lambda}_{\mu~~c}^{~~a}\hat{\Lambda}_{\nu}^{~cb}-\hat{\Lambda}_{\nu~~c}^{~~a}\hat{\Lambda}_{\mu}^{~cb}\right)\hat{e}_{a}^{~\mu}\hat{e}_{b}^{~\nu}+\frac{1}{K}ia\partial_{i}\hat{e}_{a}^{~\mu}\partial^{i}\hat{\Lambda}_{\mu}^{~ab}\delta_{b}^{\tilde{0}}+\hat{L}_{\psi}\right],
\ee
with $\hat{L}_{\psi}$ as given in eq.~\eqref{lagrangian2}.

Using Taylor series expansion of $\hat{\psi}(\hat{x}), \hat{\overline{\psi}}(\hat{x}), \hat{A}_{\mu}^{~ab}(\hat{x})$ around the commutative coordinates using eq.~\eqref{ksp-5} and keeping terms only upto first order in the deformation parameter, one can write $\hat{\psi}=\psi-ap^{0}x_{i}\partial^{i}\psi,~\hat{\overline{\psi}}=\overline{\psi}-ap^{0}x_{i}\partial^{i}\overline{\psi}$ and $\hat{A}_{\mu}^{~ab}=A_{\mu}^{~ab}-ap^{0}x_{i}\partial^{i}A_{\mu}^{~ab}$. Substituting these along with the explicit form of the Dirac derivative (see discussion after eq.~\eqref{fieldstrengthtensor}), deformed tetrads (eq.~\eqref{deformedtetrads}, eq.~\eqref{inversetetrads}) and gamma matrices (see Appendix-A) into eq.~\eqref{lagrangian2} gives,
\bea \label{lagrangian3}
\hat{L}_{\psi}&=&\frac{i}{2}\Big[\overline{\psi}\gamma^{\mu}\partial_{\mu}\psi-\partial_{\mu}\overline{\psi}\gamma^{\mu}\psi-\frac{i}{4}A_{\mu}^{~ab}\overline{\psi}[\sigma_{ab}, \gamma^{c}]_{+}\psi e_{c}^{~\mu}+ap^{0}(\partial_{i}\overline{\psi}\gamma^{i}\psi-\overline{\psi}\gamma^{i}\partial_{i}\psi)\\\nonumber
&-&ap^{0}(2x_{i}\partial^{i}\overline{\psi}\gamma^{\mu}\partial_{\mu}\psi+2\overline{\psi}\gamma^{\mu}\partial_{\mu}(x_{i}\partial^{i}\psi))+ap^{0}\Big(\frac{i}{4}A_{0}^{~ab}\overline{\psi}[\sigma_{ab}, \gamma^{\tilde{0}}]_{+}\psi e_{\tilde{0}}^{~0}-\frac{i}{4}A_{0}^{~\tilde{i}\tilde{j}}\overline{\psi}[\sigma_{\tilde{i}\tilde{j}}, \gamma^{c}]_{+}\psi e_{c}^{~0}\\\nonumber
&-&\frac{i}{4}A_{i}^{~ab}\overline{\psi}[\sigma_{ab}, \gamma^{\tilde{j}}]_{+}\psi e_{\tilde{j}}^{~i}-\frac{i}{4}A_{i}^{~\tilde{0}\tilde{i}}\overline{\psi}[\sigma_{\tilde{0}\tilde{i}}, \gamma^{\tilde{j}}]_{+}\psi e_{\tilde{j}}^{~i}-\frac{i}{4}A_{i}^{~\tilde{i}\tilde{0}}\overline{\psi}[\sigma_{\tilde{i}\tilde{0}}, \gamma^{\tilde{j}}]_{+}\psi e_{\tilde{j}}^{~i}\Big)\\\nonumber
&-&ap^{0}\left(\frac{3i}{4}A_{\mu}^{~ab}\overline{\psi}[\sigma_{ab}, \gamma^{c}]_{+}\psi e_{c}^{~\mu}+\frac{i}{4}A_{\mu}^{~ab}\overline{\psi}[\sigma_{ab}, \gamma^{c}]_{+}\psi x_{i}\partial^{i}e_{c}^{~\mu}\right)\Big]-\hat{m}\overline{\psi}\psi(1+3ap^{0}),
\eea
which is valid upto first order in the deformation parameter. Note that the indices $\tilde{0}, \tilde{i}$ etc. are for quantities in the flat space-time and $0, i$ etc. are for quantities in the curved space-time. Using this we find that the deformed action (eq.~\eqref{action3}) becomes,
\begin{multline} \label{action4}
\hat{S}=\int \abs{e}d^{4}x \Big[\frac{1}{2K}\Big(\hat{R}(\hat{\Gamma})+(\Lambda_{\mu~~c}^{~~a}\Lambda_{\nu}^{~cb}-\Lambda_{\nu~~c}^{~~a}\Lambda_{\mu}^{~cb})e_{a}^{~\mu}e_{b}^{~\nu}-ap^{0}e_{a}^{~\mu}e_{b}^{~\nu}(x_{l}\partial^{l}\Lambda_{\mu~~c}^{~~a}\Lambda_{\nu}^{~cb}-x_{l}\partial^{l}\Lambda_{\nu~~c}^{~~a}\Lambda_{\mu}^{~cb}\\+\Lambda_{\mu~~c}^{~~a}x_{l}\partial^{l}\Lambda_{\nu}^{~cb}-\Lambda_{\nu~~c}^{~~a}x_{l}\partial^{l}\Lambda_{\mu}^{~cb})+2ap^{0}(\Lambda_{\mu~~c}^{~~a}\Lambda_{i}^{~cb}e_{a}^{~\mu}e_{b}^{~i}+\Lambda_{i~~c}^{~~a}\Lambda_{\nu}^{~cb}e_{a}^{~i}e_{b}^{~\nu}-\Lambda_{\nu~~c}^{~~a}\Lambda_{i}^{~cb}e_{a}^{~i}e_{b}^{~\nu}-\Lambda_{i~~c}^{~~a}\Lambda_{\mu}^{~cb}e_{a}^{~\mu}e_{b}^{~i})\Big)\\+\frac{ia}{K}\partial_{i}e_{a}^{~\mu}\partial^{i}\Lambda_{\mu}^{~ab}\delta_{b}^{\tilde{0}}+\hat{L}_{\psi}\Big],
\end{multline}
where $\hat{L}_{\psi}$ is given in eq.~\eqref{lagrangian3}.

Varying the above action with respect to $\Lambda$ gives the equation for the deformed contortion tensor with non-commutative corrections upto first order in $ap^{0}$ as,
\bea \label{lambda}
\Lambda_{\mu}^{~ab}&=&\frac{K}{8}\overline{\psi}[\sigma^{ab}, \gamma_{c}]_{+}e_{\mu}^{~c}\psi-2ap^{0}\left(\frac{K}{8}\overline{\psi}[\sigma^{\tilde{i}b}, \gamma_{c}]_{+}e_{\mu}^{~c}\psi\delta_{\tilde{i}}^{a}-\frac{K}{8}\overline{\psi}[\sigma^{ab}, \gamma_{c}]_{+}e_{i}^{~c}\psi\delta_{\mu}^{i}\right)\\ \nonumber 
&+&ap^{0}\left(\frac{K}{8}\overline{\psi}[\sigma^{ab}, \gamma_{\tilde{j}}]_{+}e_{i}^{~\tilde{j}}\psi\delta_{\mu}^{i}+\frac{K}{8}\overline{\psi}[\sigma^{\tilde{0}\tilde{i}}, \gamma_{\tilde{j}}]_{+}e_{i}^{~\tilde{j}}\psi\delta_{\mu}^{i}\delta_{\tilde{0}}^{a}\delta_{\tilde{i}}^{b}+\frac{K}{8}\overline{\psi}[\sigma^{\tilde{i}\tilde{0}}, \gamma_{\tilde{j}}]_{+}e_{i}^{~\tilde{j}}\psi\delta_{\mu}^{i}\delta_{\tilde{i}}^{a}\delta_{\tilde{0}}^{b}\right)\\\nonumber
&-&ap^{0}\left(\frac{K}{8}\overline{\psi}[\sigma^{ab}, \gamma_{\tilde{0}}]_{+}e_{0}^{~\tilde{0}}\delta_{\mu}^{0}\psi-\frac{K}{8}\overline{\psi}[\sigma^{\tilde{i}\tilde{j}}, \gamma_{c}]_{+}e_{\mu}^{~c}\psi\delta_{\mu}^{0}\delta_{\tilde{i}}^{a}\delta_{\tilde{j}}^{b}\right)\\\nonumber
&+&ap^{0}\left(\frac{K}{8}\overline{\psi}[\sigma^{ab}, \gamma_{c}]_{+}x_{l}\partial^{l}e_{\mu}^{~c}\psi-\frac{K}{8}e_{d}^{\nu}\overline{\psi}[\sigma^{bc}, \gamma^{d}]_{+}\psi x_{l}\partial^{l}(e_{\nu}^{~a}e_{\mu c})\right)-ia\partial_{i}\partial^{i}e_{\mu}^{a}\delta_{\tilde{0}}^{b}.
\eea
Substituting the above solution back into the action in eq.~\eqref{action4} gives,
\be  \label{action5}
\hat{S}=\int \abs{e}d^{4}x\left(\frac{1}{2K}\hat{R}(\hat{\Lambda})+L_{\psi}\right),
\ee
where the Lagrangian of the fermionic field in the presence of gravity in $\kappa$-deformed space-time $(\hat{L}_{\psi})$ is,
\bea \label{lagrangian4} \nonumber
L_{\psi}&=&\frac{i}{2}\Big(\overline{\psi}\gamma^{\mu}\partial_{\mu}\psi-\partial_{\mu}\overline{\psi}\gamma^{\mu}\psi+ap^{0}\left(\partial_{i}\overline{\psi}\gamma^{i}\psi-\overline{\psi}\gamma^{i}\partial_{i}\psi\right)-ap^{0}\left(2x_{i}\partial^{i}\overline{\psi}\gamma^{\mu}\partial_{\mu}\psi+2\overline{\psi}\gamma^{\mu}\partial_{\mu}(x_{i}\partial^{i}\psi)\right)\Big)\\\nonumber
&+&\frac{1}{8}\omega_{\mu}^{~ab}\overline{\psi}[\sigma_{ab}, \gamma^{c}]_{+}e_{c}^{~\mu}\psi(1+3ap^{0})+ap^{0}\Big(-\frac{1}{8}\omega_{0}^{~ab}\overline{\psi}[\sigma_{ab}, \gamma^{\tilde{0}}]_{+}e_{\tilde{0}}^{~0}\psi+\frac{1}{8}\omega_{0}^{~\tilde{i}\tilde{j}}\overline{\psi}[\sigma_{\tilde{i}\tilde{j}}, \gamma^{d}]_{+}e_{d}^{~0}\psi\\\nonumber&+&\frac{1}{8}\omega_{i}^{~ab}\overline{\psi}[\sigma_{ab}, \gamma^{\tilde{j}}]_{+}e_{\tilde{j}}^{~i}\psi+\frac{1}{8}\omega_{i}^{~\tilde{0}\tilde{i}}\overline{\psi}[\sigma_{\tilde{0}\tilde{i}}, \gamma^{\tilde{j}}]_{+}e_{\tilde{j}}^{~i}\psi+\frac{1}{8}\omega_{i}^{~\tilde{i}\tilde{0}}\overline{\psi}[\sigma_{\tilde{i}\tilde{0}}, \gamma^{\tilde{j}}]_{+}e_{\tilde{j}}^{~i}\psi\Big)\\\nonumber
&+&ap^{0}\frac{1}{8}\omega_{\mu}^{~ab}\overline{\psi}[\sigma_{ab}, \gamma^{c}]_{+}\psi x_{l}\partial^{l}e_{c}^{~\mu}-\frac{3K}{16}\left(\overline{\psi}\gamma^{d}\gamma^{5}\psi\right)^{2}
-ap^{0}\frac{K}{8}\left(\frac{9}{2}\left(\overline{\psi}\gamma^{d}\gamma^{5}\psi\right)^{2}-\left(\overline{\psi}\gamma^{\tilde{0}}\gamma^{5}\psi\right)^{2}\right)\\\nonumber
&+&ap^{0}\frac{K}{16}\Big(-2\left(\overline{\psi}\gamma^{d}\gamma^{5}\psi\right)^{2}e_{\mu}^{~c}x_{l}\partial^{l}e_{c}^{~\mu}+\overline{\psi}\gamma^{d}\gamma^{5}\psi\overline{\psi}\gamma^{a}\gamma^{5}\psi e_{a}^{~\mu}x_{l}\partial^{l}e_{\mu d}\\\nonumber
&+&\overline{\psi}\gamma^{k}\gamma^{5}\psi\overline{\psi}\gamma_{c}\gamma^{5}\psi e_{k}^{~\mu}x_{l}\partial^{l}e_{\mu}^{~c}\Big)
+ia\Big(\frac{1}{16}\partial_{i}\partial^{i}e_{\nu}^{~a}\delta_{c}^{\tilde{0}}\overline{\psi}[\sigma^{cb}, \gamma_{a}]_{+}\psi e_{b}^{~\nu}\\
&+&\frac{1}{16}\overline{\psi}[\sigma^{a}_{~c}, \gamma_{b}]_{+}\psi\partial_{i}\partial^{i}e_{\mu}^{~c}\delta_{\tilde{0}}^{b} e_{a}^{~\mu}-\frac{1}{16}\partial_{i}\partial^{i}e_{a}^{~\mu}\delta_{b}^{\tilde{0}}\overline{\psi}[\sigma^{ab}, \gamma_{d}]_{+}\psi e_{\mu}^{~d}\Big)
-\hat{m}\overline{\psi}\psi(1+3ap^{0}).
\eea
The equation of motion of $\psi$ in the presence of gravity which is valid upto first order in $ap^{0}$ in $\kappa$-deformed space-time following from the above Lagrangian is,
\begin{multline} \label{diraceqn}
i\gamma^{\mu}\partial_{\mu}\psi(1+3ap^{0})-iap^{0}\gamma^{i}\partial_{i}\psi-\frac{i}{2}ap^{0}(2\gamma^{\mu}\partial_{\mu}(x_{i}\partial^{i}\psi)-2x^{i}\partial_{i}(\gamma^{\mu}\partial_{\mu}\psi))+\frac{1}{4}\omega_{\mu}^{~ab}\gamma^{\mu}\sigma_{ab}\psi(1+3ap^{0})\\+ap^{0}\left(-\frac{1}{4}\omega_{0}^{~ab}\gamma^{0}\sigma_{ab}\psi+\frac{1}{4}\omega_{0}^{~\tilde{i}\tilde{j}}\gamma^{0}\sigma_{\tilde{i}\tilde{j}}\psi+\frac{1}{4}\omega_{i}^{~ab}\gamma^{i}\sigma_{ab}\psi+\frac{1}{4}\omega_{i}^{~\tilde{0}\tilde{i}}\gamma^{i}\sigma_{\tilde{0}\tilde{i}}\psi+\frac{1}{4}\omega_{i}^{~\tilde{i}\tilde{0}}\gamma^{i}\sigma_{\tilde{i}\tilde{0}}\psi\right)\\+ap^{0}\frac{1}{8}\omega_{\mu}^{~ab}[\sigma_{ab}, \gamma^{c}]_{+}\psi x_{l}\partial^{l}e_{c}^{~\mu}-\frac{3K}{8}\left(\overline{\psi}\gamma^{d}\gamma^{5}\psi\right)\gamma_{d}\gamma^{5}\psi-ap^{0}\frac{K}{4}\left(\frac{9}{2}\left(\overline{\psi}\gamma^{d}\gamma^{5}\psi\right)\gamma_{d}\gamma^{5}\psi-\left(\overline{\psi}\gamma^{\tilde{0}}\gamma^{5}\psi\right)\gamma_{\tilde{0}}\gamma^{5}\psi\right)\\+ap^{0}\frac{K}{16}\Big(-4\left(\overline{\psi}\gamma^{d}\gamma^{5}\psi\right)\gamma_{d}\gamma^{5}\psi e_{\mu}^{~c}x_{l}\partial^{l}e_{c}^{~\mu}+\gamma^{d}\gamma^{5}\psi\overline{\psi}\gamma^{a}\gamma^{5}\psi e_{a}^{~\mu}x_{l}\partial^{l}e_{\mu d}+\overline{\psi}\gamma^{d}\gamma^{5}\psi\gamma^{a}\gamma^{5}\psi e_{a}^{~\mu}x_{l}\partial^{l}e_{\mu d}\\+\gamma^{d}\gamma^{5}\psi\overline{\psi}\gamma_{a}\gamma^{5}\psi e_{d}^{~\mu}x_{l}\partial^{l}e_{\mu}^{~a}+\overline{\psi}\gamma^{d}\gamma^{5}\psi\gamma_{a}\gamma^{5}\psi e_{d}^{~\mu}x_{l}\partial^{l}e_{\mu}^{~a}\Big)+ia\Big(\frac{1}{16}\partial_{i}\partial^{i}e_{\nu}^{~a}\delta_{c}^{\tilde{0}}[\sigma^{cb},\gamma_{a}]_{+}\psi e_{b}^{~\nu}\\+\frac{1}{16}[\sigma^{a}_{~c}, \gamma_{b}]_{+}\partial_{i}\partial^{i}e_{\mu}^{~c}\delta_{\tilde{0}}^{b}e_{a}^{~\mu}-\frac{1}{4}\partial_{i}\partial^{i}e_{a}^{~\mu}\delta_{b}^{\tilde{0}}[\sigma^{ab},\gamma_{d}]_{+}\psi e_{\mu}^{~d}\Big)-\hat{m}\psi(1+3ap^{0})=0.
\end{multline}
Here we have used $\frac{1}{8}\omega_{\mu}^{~ab}[\sigma_{ab}, \gamma^{c}]_{+}\psi e_{c}^{~\mu}=\frac{1}{4}\omega_{\mu}^{~ab}\gamma^{\mu}\sigma_{ab}\psi$. The above is the equation for a single species of fermions. Since fermions of all species interact with gravity through their coupling to the contortion tensor, we will now generalise eq.~\eqref{lambda} to incorporate these interactions. Further, note that each species can couple to the contortion tensor with different coupling strength\cite{amitabha}. Thus, eq.~\eqref{lambda} for all species of fermions can be written as,
\bea \label{lambda2} \nonumber
\Lambda_{\mu}^{~ab}&=&\sum_{m}\left[-\lambda_{L}^{m}\overline{\psi}_{mL}\gamma^{d}\psi_{mL}+\lambda_{R}^{m}\overline{\psi}_{mR}\gamma^{d}\psi_{mR}\right]\Big(\frac{K}{4} \epsilon^{ab}_{~cd}e_{\mu}^{~c}-ap^{0}\frac{K}{2}\Big(\epsilon^{\tilde{i}b}_{~cd}e_{\mu}^{~c}\delta_{\tilde{i}}^{a}-\epsilon^{ab}_{~cd}e_{i}^{~c}\delta_{\mu}^{i}\Big)\\\nonumber
&+&ap^{0}\frac{K}{4}\Big(-\epsilon^{ab}_{~\tilde{0}d}e_{0}^{~\tilde{0}}\delta_{\mu}^{0}+\epsilon^{\tilde{i}\tilde{j}}_{~cd}e_{0}^{~c}\delta_{\mu}^{0}\delta_{\tilde{i}}^{a}\delta_{\tilde{j}}^{b}+\epsilon^{ab}_{~\tilde{j}d}e_{i}^{~\tilde{j}}\delta_{\mu}^{i}
+\epsilon^{\tilde{0}\tilde{i}}_{~\tilde{j}d}e_{i}^{~\tilde{j}}\delta_{\mu}^{i}\delta_{\tilde{0}}^{a}\delta_{\tilde{i}}^{b}+\epsilon^{\tilde{i}\tilde{0}}_{~\tilde{j}d}e_{i}^{~\tilde{j}}\delta_{\mu}^{i}\delta_{\tilde{i}}^{a}\delta_{\tilde{0}}^{b}\Big)\\
&+&ap^{0}\frac{K}{4}\Big(\epsilon^{ab}_{~cd}x_{l}\partial^{l}e_{\mu}^{~c}-\epsilon^{bck}_{~~d}e_{k}^{~\nu}x_{l}\partial^{l}(e_{\nu}^{~a}e_{\mu c})\Big)\Big)-ia\partial_{i}\partial^{i}e_{\mu}^{~a}\delta_{\tilde{0}}^{b}.
\eea
In obtaining the above equation, $[\sigma_{ab}, \gamma_{c}]_{+}=2\epsilon_{abcd}\gamma^{d}\gamma^{5},~\overline{\psi}_{mL}\gamma^{d}\gamma^{5}\psi_{mL}=-\overline{\psi}_{mL}\gamma^{d}\psi_{mL}$ and $\overline{\psi}_{mR}\gamma^{d}\gamma^{5}\psi_{mR}=\overline{\psi}_{mR}\gamma^{d}\psi_{mR}$ have been used. Substituting the above solution back into the action for all species of fermions gives the Lagrangian valid upto first order in the deformation parameter as,
\bea \label{lagrangian5} 
\hat{L}_{\psi}&=&\frac{i}{2}\Big(\overline{\psi}_{m}\gamma^{\mu}\partial_{\mu}\psi_{m}-\partial_{\mu}\overline{\psi}_{m}\gamma^{\mu}\psi_{m}+ap^{0}\left(\partial_{j}\overline{\psi}_{m}\gamma^{j}\psi_{m}-\overline{\psi}_{m}\gamma^{j}\partial_{j}\psi_{m}\right)\\\nonumber
&-&ap^{0}\left(2x_{j}\partial^{j}\overline{\psi}_{m}\gamma^{\mu}\partial_{\mu}\psi_{m}+2\overline{\psi}_{m}\gamma^{\mu}\partial_{\mu}(x_{j}\partial^{j}\psi_{m})\right)\Big)
+\frac{1}{8}\omega_{\mu}^{~ab}\overline{\psi}_{m}[\sigma_{ab}, \gamma^{c}]_{+}e_{c}^{~\mu}\psi_{m}(1+3ap^{0})\\\nonumber
&+&ap^{0}\Big(-\frac{1}{8}\omega_{0}^{~ab}\overline{\psi}_{m}[\sigma_{ab}, \gamma^{\tilde{0}}]_{+}e_{\tilde{0}}^{~0}\psi_{m}+\frac{1}{8}\omega_{0}^{~\tilde{i}\tilde{j}}\overline{\psi}_{m}[\sigma_{\tilde{i}\tilde{j}}, \gamma^{d}]_{+}e_{d}^{~0}\psi_{m}
+\frac{1}{8}\omega_{k}^{~ab}\overline{\psi}_{m}[\sigma_{ab}, \gamma^{\tilde{j}}]_{+}e_{\tilde{j}}^{~k}\psi_{m}\\\nonumber
&+&\frac{1}{8}\omega_{k}^{~\tilde{0}\tilde{i}}\overline{\psi}_{m}[\sigma_{\tilde{0}\tilde{i}}, \gamma^{\tilde{j}}]_{+}e_{\tilde{j}}^{~k}\psi_{i}+\frac{1}{8}\omega_{k}^{~\tilde{i}\tilde{0}}\overline{\psi}_{m}[\sigma_{\tilde{i}\tilde{0}}, \gamma^{\tilde{j}}]_{+}e_{\tilde{j}}^{~k}\psi_{m}\Big)+ap^{0}\frac{1}{8}\omega_{\mu}^{~ab}\overline{\psi}_{m}[\sigma_{ab}, \gamma^{c}]_{+}\psi_{m} x_{l}\partial^{l}e_{c}^{~\mu}\\\nonumber
&-&\frac{1}{2}\left[\lambda_{m}^{V}\overline{\psi}_{m}\gamma^{d}\psi_{m}+\lambda_{m}^{A}\overline{\psi}_{m}\gamma^{d}\gamma^{5}\psi_{m}\right]^{2}(1+3ap^{0})
+ap^{0}\frac{1}{3}\left[\lambda_{m}^{V}\overline{\psi}_{m}\gamma^{\tilde{0}}\psi_{m}+\lambda_{m}^{A}\overline{\psi}_{m}\gamma^{\tilde{0}}\gamma^{5}\psi_{m}\right]^{2}\\\nonumber
&+&ap^{0}\Big(-\frac{1}{3}e_{\mu}^{~c}x_{l}\partial^{l}e_{c}^{~\mu}\left[\lambda_{m}^{V}\overline{\psi}_{m}\gamma^{d}\psi_{m}+\lambda_{m}^{A}\overline{\psi}_{m}\gamma^{d}\gamma^{5}\psi_{m}\right]^{2}\\\nonumber
&+&\frac{1}{3}e_{a}^{~\mu}x_{l}\partial^{l}e_{\mu d}\left[\lambda_{m}^{V}\overline{\psi}_{m}\gamma^{d}\psi_{m}+\lambda_{m}^{A}\overline{\psi}_{m}\gamma^{d}\gamma^{5}\psi_{m}\right]\left[\lambda_{m}^{V}\overline{\psi}_{m}\gamma^{a}\psi_{m}+\lambda_{m}^{A}\overline{\psi}_{m}\gamma^{a}\gamma^{5}\psi_{m}\right]\Big)\\\nonumber
&+&ia\left[\lambda_{m}^{V}\overline{\psi}_{m}\gamma^{d}\psi_{m}+\lambda_{m}^{A}\overline{\psi}_{m}\gamma^{d}\gamma^{5}\psi_{m}\right]\Big(\partial_{m}\partial^{m}e_{\mu}^{~a}\delta_{c}^{\tilde{0}}e_{b}^{~\mu}\frac{1}{8}\epsilon^{cb}_{~ad}+\frac{1}{8}\epsilon^{a}_{~cbd}e_{a}^{~\mu}\partial_{i}\partial^{i}e_{\mu}^{~c}\delta_{\tilde{0}}^{b}\\\nonumber
&-&\partial_{i}\partial^{i}e_{a}^{~\mu}\delta_{b}^{\tilde{0}}e_{\mu}^{~c}\frac{1}{2}\epsilon^{ab}_{~cd}\Big)-\hat{m}\overline{\psi}\psi(1+3ap^{0}).
\eea
In writing the above equation,the index $m$ is for representing all the species of fermions which is summed over, a factor $\sqrt{\frac{3K}{8}}$ appearing in the quartic interaction terms has been absorbed into $\lambda_{m}^{V,A}$ where, 
\be \label{couplingconstant}
\lambda_{m}^{A}=\frac{1}{2}(\lambda_{R}+\lambda_{L}),~\lambda_{m}^{V}=\frac{1}{2}(\lambda_{R}-\lambda_{L}).
\ee
Hence, the Dirac equation for all species of fermions in the presence of gravity in $\kappa$-deformed space-time following from the above Lagrangian is,
\begin{multline} \label{diraceqn2}
i\gamma^{\mu}\partial_{\mu}\psi_{m}(1+3ap^{0})-iap^{0}\gamma^{j}\partial_{j}\psi_{m}-\frac{i}{2}ap^{0}(2\gamma^{\mu}\partial_{\mu}(x_{j}\partial^{j}\psi_{m})-2x^{j}\partial_{j}(\gamma^{\mu}\partial_{\mu}\psi_{m}))+\frac{1}{4}\omega_{\mu}^{~ab}\gamma^{\mu}\sigma_{ab}\psi_{m}(1+3ap^{0})\\+ap^{0}\left(-\frac{1}{4}\omega_{0}^{~ab}\gamma^{0}\sigma_{ab}\psi_{m}+\frac{1}{4}\omega_{0}^{~\tilde{i}\tilde{j}}\gamma^{0}\sigma_{\tilde{i}\tilde{j}}\psi_{m}+\frac{1}{4}\omega_{j}^{~ab}\gamma^{j}\sigma_{ab}\psi_{m}+\frac{1}{4}\omega_{j}^{~\tilde{0}\tilde{i}}\gamma^{j}\sigma_{\tilde{0}\tilde{i}}\psi_{m}+\frac{1}{4}\omega_{j}^{~\tilde{i}\tilde{0}}\gamma^{j}\sigma_{\tilde{i}\tilde{0}}\psi_{m}\right)\\+ap^{0}\frac{1}{8}\omega_{\mu}^{~ab}[\sigma_{ab}, \gamma^{c}]_{+}\psi_{m} x_{l}\partial^{l}e_{c}^{~\mu}-\left[\sum_{f}\lambda_{f}^{V}\overline{\psi}_{f}\gamma^{d}\psi_{f}+\lambda_{f}^{A}\overline{\psi}_{f}\gamma^{d}\gamma^{5}\psi_{f}\right]\left[\lambda_{m}^{V}\gamma_{d}\psi_{m}+\lambda_{m}^{A}\gamma_{d}\gamma^{5}\psi_{m}\right](1+3ap^{0})\\+ap^{0}\frac{2}{3}\left[\sum_{f}\lambda_{f}^{V}\overline{\psi}_{f}\gamma^{\tilde{0}}\psi_{f}+\lambda_{f}^{A}\overline{\psi}_{f}\gamma^{\tilde{0}}\gamma^{5}\psi_{f}\right]\left[\lambda_{m}^{V}\gamma_{\tilde{0}}\psi_{m}+\lambda_{m}^{A}\gamma_{\tilde{0}}\gamma^{5}\psi_{m}\right]\\+ap^{0}\Big(-\frac{2}{3}e_{\mu}^{~c}x_{l}\partial^{l}e_{c}^{~\mu}\left[\sum_{f}\lambda_{f}^{V}\overline{\psi}_{f}\gamma^{d}\psi_{f}+\lambda_{f}^{A}\overline{\psi}_{f}\gamma^{d}\gamma^{5}\psi_{f}\right]\left[\lambda_{m}^{V}\gamma_{d}\psi_{m}+\lambda_{m}^{A}\gamma_{d}\gamma^{5}\psi_{m}\right]\Big)\\+ia\left[\lambda_{m}^{V}\gamma^{d}\psi_{m}+\lambda_{m}^{A}\gamma^{d}\gamma^{5}\psi_{m}\right]\Big(\frac{1}{8}\partial_{i}\partial^{i}e_{\mu}^{~a}\delta_{c}^{\tilde{0}}e_{b}^{~\mu}\epsilon^{cb}_{~ad}+\frac{1}{8}\epsilon^{a}_{~cbd}e_{a}^{~\mu}\partial_{i}\partial^{i}e_{\mu}^{~c}\delta_{\tilde{0}}^{b}-\frac{1}{2}\partial_{i}\partial^{i}e_{a}^{~\mu}\delta_{b}^{\tilde{0}}e_{\mu}^{~c}\epsilon^{ab}_{~cd}\Big)\\-\hat{m}\psi(1+3ap^{0})=0.
\end{multline}
In the above, $\psi_{m}$ stands for the neutrino fields we are studying while $\psi_{f}$ are the fermionic fields representing all the background fermionic matter. Assuming that there are no other sources of gravity affecting fermions moving through the background matter, we can neglect $\omega_{\mu}^{~ab}$. Since the interaction is weak, the terms in the square bracket in eq.~\eqref{diraceqn2} for $f \neq m$ are replaced by their averages, i.e., $\left[\sum_{f}\lambda_{f}^{V}\overline{\psi}_{f}\gamma^{d}\psi_{f}+\lambda_{f}^{A}\overline{\psi}_{f}\gamma^{d}\gamma^{5}\psi_{f}\right]=\sum\limits_{f \neq i}\left<\lambda_{f}^{V}\overline{\psi}_{f}\gamma^{d}\psi_{f}+\lambda_{f}^{A}\overline{\psi}_{f}\gamma^{d}\gamma^{5}\psi_{f}\right>$ for fermions passing through ordinary matter. Now, if the background matter is at rest, the average of spin density $\left(\left<\overline{\psi}_{f}\gamma^{d}\gamma^{5}\psi_{f}\right>\right)$ and the average of momentum density $\left(\left<\overline{\psi}_{f}\gamma^{i}\psi_{f}\right>\right)$ go to zero whereas $\left<\overline{\psi}_{f}\gamma^{\tilde{0}}\psi_{f}\right>=\tilde{n}_{\lambda}^{i}$ becomes the number density of fermions of type $f$\cite{amitabha}. With these considerations, the above Dirac equation reduces to,
\begin{multline} \label{diraceqn3}
i\gamma^{\mu}\partial_{\mu}\psi_{m}(1+3ap^{0})-iap^{0}\gamma^{j}\partial_{j}\psi_{m}-\frac{i}{2}ap^{0}(2\gamma^{\mu}\partial_{\mu}(x_{j}\partial^{j}\psi_{m})-2x^{j}\partial_{j}(\gamma^{\mu}\partial_{\mu}\psi_{m}))\\-\tilde{n}_{\lambda}^{m}\left[\lambda_{m}^{V}\gamma_{\tilde{0}}\psi_{m}+\lambda_{m}^{A}\gamma_{\tilde{0}}\gamma^{5}\psi_{m}\right]\left(1+\frac{7}{3}ap^{0}\right)+ap^{0}\tilde{n}_{\lambda}^{m}\left[\lambda_{m}^{V}\gamma_{\tilde{0}}\psi_{m}+\lambda_{m}^{A}\gamma_{\tilde{0}}\gamma^{5}\psi_{m}\right]\Big(-\frac{2}{3}e_{\mu}^{~c}x_{l}\partial^{l}e_{c}^{~\mu}+\frac{2}{3}e_{\tilde{0}}^{~\mu}x_{l}\partial^{l}e_{\mu \tilde{0}}\Big)\\
+ia\left[\lambda_{m}^{V}\gamma^{d}\psi_{m}+\lambda_{m}^{A}\gamma^{d}\gamma^{5}\psi_{m}\right]\Big(\frac{1}{8}\partial_{i}\partial^{i}e_{\mu}^{~a}\delta_{c}^{\tilde{0}}e_{b}^{~\mu}\epsilon^{cb}_{~ad}+\frac{1}{8}\epsilon^{a}_{~cbd}e_{a}^{~\mu}\partial_{i}\partial^{i}e_{\mu}^{~c}\delta_{\tilde{0}}^{b}-\frac{1}{2}\partial_{i}\partial^{i}e_{a}^{~\mu}\delta_{b}^{\tilde{0}}e_{\mu}^{~c}\epsilon^{ab}_{~cd}\Big)\\-\hat{m}\psi(1+3ap^{0})=0.
\end{multline}
Note that in the limit $a \rightarrow 0$, we obtain the commutative Dirac equation.

Now, in order to see the effect of gravity on propagating fermions in $\kappa$-deformed space-time, we consider the effective quartic interaction term appearing in the Lagrangian given in eq.~\eqref{lagrangian5} which is,
\begin{multline} \label{interaction1}
-\frac{1}{2}\left[\lambda_{m}^{V}\overline{\psi}_{m}\gamma^{d}\psi_{m}+\lambda_{m}^{A}\overline{\psi}_{m}\gamma^{d}\gamma^{5}\psi_{m}\right]^{2}(1+3ap^{0})+ap^{0}\frac{1}{3}\left[\lambda_{m}^{V}\overline{\psi}_{m}\gamma^{\tilde{0}}\psi_{m}+\lambda_{m}^{A}\overline{\psi}_{m}\gamma^{\tilde{0}}\gamma^{5}\psi_{m}\right]^{2}\\+ap^{0}\Big(-\frac{1}{3}e_{\mu}^{~c}x_{l}\partial^{l}e_{c}^{~\mu}\left[\lambda_{m}^{V}\overline{\psi}_{m}\gamma^{d}\psi_{m}+\lambda_{m}^{A}\overline{\psi}_{m}\gamma^{d}\gamma^{5}\psi_{m}\right]^{2}\\+\frac{1}{3}e_{a}^{~\mu}x_{l}\partial^{l}e_{\mu d}\left[\lambda_{m}^{V}\overline{\psi}_{m}\gamma^{d}\psi_{m}+\lambda_{m}^{A}\overline{\psi}_{m}\gamma^{d}\gamma^{5}\psi_{m}\right]\left[\lambda_{m}^{V}\overline{\psi}_{m}\gamma^{a}\psi_{m}+\lambda_{m}^{A}\overline{\psi}_{m}\gamma^{a}\gamma^{5}\psi_{m}\right]\Big).
\end{multline}
We see that the non-commutative correction in the last two terms of the effective quartic interaction term have explicit dependence on the background metric through the tetrads, whereas the commutative part of the interaction term has no explicit dependence on the background metric \cite{amitabha}. To see how this interaction term will affect the effective mass of the propagating fermions, we simplify the above equation using the aforementioned approximations (see discussions after eq.~\eqref{diraceqn2}) which gives,
\be \label{interaction2}
-\frac{1}{2}\tilde{n}\left[\lambda_{m}^{v}\overline{\psi}_{m}\gamma^{\tilde{0}}\psi_{m}+\lambda_{m}^{A}\overline{\psi}_{m}\gamma^{\tilde{0}}\gamma^{5}\psi_{m}\right]\left(1+\frac{7}{3}ap^{0}+ap^{0}\left(\frac{2}{3}e_{\mu}^{~c}x_{l}\partial^{l}e_{c}^{~\mu}-\frac{2}{3}e_{\tilde{0}}^{~\mu}x_{l}\partial^{l}e_{\mu}^{~\tilde{0}}\right)\right).
\ee
The term $\tilde{n}$ represents the number density of the background matter, earlier denoted as $\tilde{n}_{\lambda}^{m}$ and it is seen to contribute to the effective mass of the propagating fermion. The above term is the effective quartic interaction term appearing in the Lagrangian because of the effect of gravity on fermions propagating in deformed space-time. We are interested to see how the effective mass of a propagating neutrino is affected by gravity in $\kappa$-deformed space-time. The effect of curvature on neutrino oscillation cannot be neglected as the gravitational interaction is dependent on the coupling constants which have to be fixed by experiments \cite{amitabha}. We use eq.~\eqref{interaction2} to see how the inclusion of gravitational interactions of neutrinos affects the  neutrino oscillation probabilities.

\section{Neutrino Oscillations in $\kappa$-deformed space-time: Two flavour} 

In this section, we derive the oscillation probabilities for two flavours of neutrinos in vacuum and in the presence of gravitational interactions for neutrinos moving through constant matter density in $\kappa$-deformed space-time.

\subsection{Two-flavour oscillation in vacuum}
The free particle dispersion relation in $\kappa$-deformed space-time is given as \cite{meljanac},
\be \label{dispnrelatn}
\frac{4}{a^{2}}sinh^{2}\left(\frac{A}{2}\right)-p_{i}p_{i}\frac{e^{-A}}{\varphi^{2}(A)}-m^{2}c^{2}+\frac{a^{2}}{4}\left[\frac{4}{a^{2}}sinh^{2}\left(\frac{A}{2}\right)-p_{i}p_{i}\frac{e^{-A}}{\varphi^{2}(A)}\right]^{2}=0. 
\ee
where we choose $\varphi(ap^{0})=e^{-ap^{0}}$ (see the discussion in Section $2$). Keeping upto first order in the deformation parameter $ap^{0}$, the dispersion relation reduces to,
\be \label{dispnrelatn2}
\hat{E}=E\left(1+\frac{ap^{0}}{2}\left(1-\frac{m^2}{E^{2}}\right)\right).
\ee
Here $E^{2}=p^{2}+m^{2}$, and in the limit $a\rightarrow 0$, we get back the commutative dispersion relation.
Using this, we find the energy eigenvalues of mass eigenstates $\nu_{1}$ and $\nu_{2}$ of neutrinos as,
\be \label{energyeigenvalues}
\hat{E}_{i}=E\left(1+\frac{ap^{0}}{2}\left(\frac{1}{1+\frac{m_{i}^{2}}{E^2}}\right)\right)+\frac{m_{i}^{2}}{2E}\left(1+\frac{ap^{0}}{2}\left(\frac{1}{1+\frac{m_{i}^{2}}{E^2}}\right)\right),
\ee
where $i=1, 2$ and all massive neutrinos are assumed to have the same momentum \cite{monojith}. The time dependent Schr$\ddot{o}$dinger equation in mass basis in $\kappa$-deformed space-time is written as,
\be \label{schrodingereqn1}
i\frac{\partial \nu_{i}}{\partial t}=\hat{H}_{M}\nu_{i},~~~i=1,2.
\ee
where $\hat{H}_{M}$ is the deformed effective Hamiltonian in the mass basis with the mass eigenstate $\nu_{i}$. The deformed effective Hamiltonian for two flavours of neutrinos is given as $diag(\hat{E}_{1}, \hat{E}_{2})$. In the remaining part of the calculations, we neglect the common diagonal terms as they are the same for all neutrinos and doesn't affect the neutrino oscillation probability. The explicit form of the deformed effective Hamiltonian in mass basis after subtracting the additive common factors from the diagonal elements is,
\be \label{hamiltonian1}
\hat{H}_{M}=\begin{bmatrix}
\frac{m_{1}^{2}}{2E}\left(1+\frac{ap^{0}}{2}\left(\frac{1}{1+\frac{m_{1}^{2}}{E^{2}}}\right)\right)+\frac{ap^{0}}{2}\frac{E}{1+\frac{m_{1}^{2}}{E^2}} & 0\\
0 & \frac{m_{2}^{2}}{2E}\left(1+\frac{ap^{0}}{2}\left(\frac{1}{1+\frac{m_{2}^{2}}{E^{2}}}\right)\right)+\frac{ap^{0}}{2}\frac{E}{1+\frac{m_{2}^{2}}{E^2}}
\end{bmatrix}.
\ee
The neutrinos in the mass basis are related to those in the flavour basis through a $2\times 2$ unitary mixing matrix $U$ that is parametrised by mixing angle $\theta$ as,
\be 
\begin{pmatrix}
\nu_{e}\\
\nu_{\mu}
\end{pmatrix}=U\begin{pmatrix}
\nu_{1}\\
\nu_{2}
\end{pmatrix}=\begin{bmatrix}
cos\theta & sin\theta\\
-sin\theta & cos\theta
\end{bmatrix}\begin{pmatrix}
\nu_{1}\\
\nu_{2}
\end{pmatrix}.
\ee
Thus, converting the Schrodinger equation (eq.~\eqref{schrodingereqn1}) to flavour basis gives the deformed effective Hamiltonian in flavour basis as $\hat{H}_{F}=U\hat{H}_{M}U^{\dagger}$. The explicit form of $\hat{H}_{F}$ after subtracting $\frac{m_{1}^{2}+m_{2}^{2}}{4E}+\frac{ap^{0}}{2}\left(\frac{m_{1}^{2}}{4E}\left(\frac{1}{1+\frac{m_{1}^{2}}{E^{2}}}\right)+\frac{m_{2}^{2}}{4E}\left(\frac{1}{1+\frac{m_{2}^{2}}{E^{2}}}\right)\right)+\frac{ap^{0}}{2}\frac{E}{\left(1+\frac{m_{1}^{2}}{E^{2}}\right)\left(1+\frac{m_{2}^{2}}{E^{2}}\right)}\left(1+\frac{m_{1}^{2}}{2E^{2}}+\frac{m_{2}^{2}}{2E^{2}}\right)$ from the diagonal terms is,
\be \label{hamiltonianF}
\hat{H}_{F}=\begin{bmatrix}
-\hat{p} & \hat{q}\\
\hat{q} & \hat{p}
\end{bmatrix},
\ee
where $\hat{p}=\frac{\Delta_{21}}{4E}\left(cos2\theta-\frac{ap^{0}}{2}\frac{cos2\theta}{\left(1+\frac{m_{1}^{2}}{E^{2}}\right)\left(1+\frac{m_{2}^{2}}{E^{2}}\right)}\right)$, $\hat{q}=\frac{\Delta_{21}}{4E}\left(sin2\theta-\frac{ap^{0}}{2}\frac{sin2\theta}{\left(1+\frac{m_{1}^{2}}{E^{2}}\right)\left(1+\frac{m_{2}^{2}}{E^{2}}\right)}\right)$ and $\Delta_{21}=m_{2}^{2}-m_{1}^{2}$. Now, transforming eq.~\eqref{schrodingereqn1} to the flavour basis, we can write,
\bea \label{coupledDE}
i\frac{d \nu_{e}}{dt}&=&-\hat{p}\nu_{e}+\hat{q}\nu_{\mu}, \\\nonumber
i\frac{d \nu_{\mu}}{dt}&=&\hat{q}\nu_{e}+\hat{p}\nu_{\mu}.
\eea
The solution for the above coupled differential equation with $\abs{\nu_{e}(t)}^{2}+\abs{\nu_{\mu}(t)}^{2}=1,~\nu_{e}(0)=1$ and $\nu_{\mu}(0)=0$ are,
\bea 
\nu_{e}(t)&=&sin^{2}\theta e^{-i\hat{\Omega}t}+cos^{2}\theta e^{i\hat{\Omega}t},\\
\nu_{\mu}(t)&=&sin\theta cos\theta e^{-i\hat{\Omega}t}-sin\theta cos\theta e^{i\hat{\Omega}t},
\eea
where
\be \label{Omega}
\hat{\Omega}=\sqrt{\hat{p}^{2}+\hat{q}^{2}}=\pm\frac{\Delta_{21}}{4E}\left(1-\frac{ap^{0}}{2}\frac{1}{\left(1+\frac{m_{1}^{2}}{E^{2}}\right)\left(1+\frac{m_{2}^{2}}{E^{2}}\right)}\right).
\ee
Using this, we find the transition probability of $\nu_{e} \rightarrow \nu_{\mu}$ in $\kappa$-deformed space-time as,
\bea \label{pemu1} 
P_{e\mu}&=&\abs{\left<\nu_{e}|\nu_{\mu}(t)\right>}^{2} \\ \nonumber
&=&sin^{2}2\theta sin^{2}\left(\frac{\Delta_{21}L}{4E}\left(1-\frac{ap^{0}}{2}\frac{1}{\left(1+\frac{m_{1}^{2}}{E^{2}}\right)\left(1+\frac{m_{2}^{2}}{E^{2}}\right)}\right)\right).
\eea
and the survival probability for $\nu_{e} \rightarrow \nu_{e}$ is
\bea \label{pee1} 
P_{ee}&=&\abs{\left<\nu_{e}|\nu_{e}(t)\right>}^{2} \\ \nonumber
&=&1-sin^{2}2\theta sin^{2}\left(\frac{\Delta_{21}L}{4E}\left(1-\frac{ap^{0}}{2}\frac{1}{\left(1+\frac{m_{1}^{2}}{E^{2}}\right)\left(1+\frac{m_{2}^{2}}{E^{2}}\right)}\right)\right).
\eea
In the limit $a \rightarrow 0$, we get back the commutative results \cite{monojith}. The non-commutative corrections to the transition probabilities are seen to depend on the mass of the neutrinos and its energy. Note that for terrestrial neutrinos, the energy of the neutrinos is in $GeV$ scale while its mass is in $eV$ scale and hence the coefficient of $ap^{0}$ dependent terms reduces to $\frac{1}{2}$, i.e., $\left(1+\frac{m_{1}^{2}}{E^{2}}\right)\left(1+\frac{m_{2}^{2}}{E^{2}}\right)\approx 1$.

\subsection{Two-flavour oscillation in the presence of matter and gravity}
On including the effects of matter, the effective Hamiltonian in the flavour basis in $\kappa$-deformed space-time becomes,
\be \label{hamiltonianmatterF}
\hat{H}_{F}=\begin{bmatrix}
-\hat{\mathcal{D}} & \hat{\mathcal{C}}\\
\hat{\mathcal{C}} & \hat{\mathcal{D}}
\end{bmatrix},
\ee
where $\hat{\mathcal{D}}=\frac{\Delta_{21}}{4E}\left(cos2\theta-\frac{ap^{0}}{2}\frac{cos2\theta}{\left(1+\frac{m_{1}^{2}}{E^{2}}\right)\left(1+\frac{m_{2}^{2}}{E^{2}}\right)}\right)-\frac{A}{4E}$, $\hat{\mathcal{C}}=\frac{\Delta_{21}}{4E}\left(sin2\theta-\frac{ap^{0}}{2}\frac{sin2\theta}{\left(1+\frac{m_{1}^{2}}{E^{2}}\right)\left(1+\frac{m_{2}^{2}}{E^{2}}\right)}\right)$ and $A=2\sqrt{2}G_{F}N_{e}E,~N_{e}$ is the electron number density of the medium and $G_{F}$ is the Fermi constant. Note that the off-diagonal elements remain the same as those of the vacuum Hamiltonian (eq.~\eqref{hamiltonianF}) but $\frac{A}{4E}$ is subtracted from the diagonal elements. 

The effective quartic interaction term due to gravity in $\kappa$-deformed space-time in the Lagrangian is as given in eq.~\eqref{interaction2}. The torsion coupling constant is assumed to be negligible for right-handed neutrinos compared to that of left-handed neutrinos\cite{amitabha} which gives the contribution of gravity to the effective Hamiltonian in the mass basis as
\be \label{gravityhamiltonian1}
\sum_{i=1,2}\left(\lambda_{i}\nu_{i}^{\dagger}P_{L}\nu_{i}\right)\tilde{n}\left(1+\frac{7}{3}ap^{0}+ap^{0}\left(\frac{2}{3}e_{\mu}^{~c}x_{l}\partial^{l}e_{c}^{~\mu}-\frac{2}{3}e_{\tilde{0}}^{~\mu}x_{l}\partial^{l}e_{\mu}^{~\tilde{0}}\right)\right),
\ee
where $P_{L}=\frac{1}{2}(1-\gamma^{5})$ is the left handed projection operator. Converting this interaction Hamiltonian in deformed space-time to the flavour basis through the mixing matrix($U$) gives the interaction Hamiltonian in the flavour basis as,
\begin{multline} \label{gravityhamiltonian2}
\hat{H}_{F}^{torsion}=\left(1+\frac{7}{3}ap^{0}+ap^{0}\left(\frac{2}{3}e_{\mu}^{~c}x_{l}\partial^{l}e_{c}^{~\mu}-\frac{2}{3}e_{\tilde{0}}^{~\mu}x_{l}\partial^{l}e_{\mu}^{~\tilde{0}}\right)\right)\tilde{n}\begin{bmatrix}
\nu_{e}^{\dagger} & \nu_{\mu}^{\dagger}
\end{bmatrix} P_{L} \\\begin{bmatrix}
\lambda_{1}cos^{2}\theta+\lambda_{2}sin^{2}\theta & (\lambda_{2}-\lambda_{1})sin\theta cos\theta\\
(\lambda_{2}-\lambda_{1})sin\theta cos\theta & \lambda_{1}sin^{2}\theta+\lambda_{2}cos^{2}\theta
\end{bmatrix} \begin{bmatrix}
\nu_{e} \\ \nu_{\mu}
\end{bmatrix}.
\end{multline} 
Now, the total effective Hamiltonian in flavour basis after subtracting $(\lambda_{1}+\lambda_{2})\frac{\tilde{n}}{2}+ap^{0}(\lambda_{1}+\lambda_{2})\frac{\tilde{n}}{2}\left(\frac{7}{3}ap^{0}+ap^{0}\left(\frac{2}{3}e_{\mu}^{~c}x_{l}\partial^{l}e_{c}^{~\mu}-\frac{2}{3}e_{\tilde{0}}^{~\mu}x_{l}\partial^{l}e_{\mu}^{~\tilde{0}}\right)\right)$ from the diagonal elements of $\hat{H}_{F}^{torsion}$ in eq.~\eqref{gravityhamiltonian2} become,
\be \label{hamiltoniangravity}
\hat{H}_{F}=\frac{1}{4E}\begin{bmatrix}
A-\Delta\hat{m}_{21}^{2}cos2\theta & \Delta\hat{m}_{21}^{2}sin2\theta\\
\Delta\hat{m}_{21}^{2}sin2\theta & -A+\Delta\hat{m}_{21}^{2}cos2\theta 
\end{bmatrix},
\ee
where 
\begin{multline} \label{massdeformation}
\Delta\hat{m}_{21}=\Delta \tilde{m}_{21}^{2}+ap^{0}\left(\frac{-\Delta_{21}}{2\left(1+\frac{m_{1}^{2}}{E^{2}}\right)\left(1+\frac{m_{2}^{2}}{E^{2}}\right)}+\frac{2}{3}\tilde{n}\Delta\lambda_{21} E\left(7+2e_{\mu}^{~c}x_{l}\partial^{l}e_{c}^{~\mu}-2e_{\tilde{0}}^{~\mu}x_{l}\partial^{l}e_{\mu}^{~\tilde{0}}\right)\right).
\end{multline}
Here, $\Delta \tilde{m}_{21}^{2}=\Delta_{21}+2\Delta\lambda_{21}\tilde{n}E$ and $\Delta\lambda_{21}=\lambda_{2}-\lambda_{1}$. The energy eigenvalue of the effective Hamiltonian (eq.~\eqref{hamiltoniangravity}) shifted by an overall factor of $\frac{A}{4E}$ is,
\be \label{energyeigenvalue2} 
\hat{E}_{1,2}=\frac{1}{4E}\left(A \pm \sqrt{\left(-A+\Delta\hat{m}_{21}^{2}cos2\theta\right)^{2}+\left(\Delta\hat{m}_{21}^{2}sin2\theta\right)^{2}}\right).
\ee
Equating the above with the difference in energy of the neutrinos obtained from the dispersion relation in eq.~\eqref{energyeigenvalues}, we get, 
\be \label{massdeformation2} 
\hat{\Delta}_{21}^{M\prime}=\hat{\Delta}_{21}^{M}\left(1+\frac{ap^{0}}{2\left(1+\frac{m_{1}^{2}}{E^{2}}\right)\left(1+\frac{m_{2}^{2}}{E^{2}}\right)}\right),
\ee
where $\hat{\Delta}_{21}^{M}=\sqrt{\left(-A+\Delta\hat{m}_{21}^{2}cos2\theta\right)^{2}+\left(\Delta\hat{m}_{21}^{2}sin2\theta\right)^{2}}$.
To see the modification of the mixing angle, the deformed effective Hamiltonian in the flavour basis is converted to the mass basis as, 
\bea 
\hat{H}_{eff}^{M}&=&U_{M}^{\dagger}\hat{H}_{eff}^{F}U_{M}, \\
&=&\frac{1}{4E}\begin{bmatrix}
cos\theta_{M} & -sin\theta_{M} \\
sin\theta_{M} & cos\theta_{M}
\end{bmatrix}\begin{bmatrix}
A-\Delta\hat{m}_{s}^{2}cos2\theta & \Delta\hat{m}_{s}^{2}sin2\theta\\
\Delta\hat{m}_{s}^{2}sin2\theta & -A+\Delta\hat{m}_{s}^{2}cos2\theta 
\end{bmatrix}\begin{bmatrix}
cos\theta_{M} & sin\theta_{M} \\
-sin\theta_{M} & cos\theta_{M}
\end{bmatrix}.
\eea
Setting the off-diagonal elements of $\hat{H}_{eff}^{M}$ to zero as the matter Hamiltonian has only diagonal elements, gives,
\be \label{mixingangle}
tan\theta_{M}=tan2\theta\left(\frac{1}{1-\frac{A}{\Delta\hat{m}_{s}^{2}cos2\theta}}\right).
\ee
Hence, the neutrino survival probability given in eq.~\eqref{pee1} in presence of matter and gravity in $\kappa$-deformed space-time becomes,
\bea \label{Pee2matter}
P_{ee}&=&1-sin^{2}2\theta_{M} sin^{2}\left(\frac{\hat{\Delta}_{21}^{M\prime}L}{4E}\left(1-\frac{ap^{0}}{2}\frac{1}{\left(1+\frac{m_{1}^{2}}{E^{2}}\right)\left(1+\frac{m_{2}^{2}}{E^{2}}\right)}\right)\right),\\
&=&1-sin^{2}2\theta_{M} sin^{2}\left(\frac{\hat{\Delta}_{21}^{M}L}{4E}\right).
\eea 
Note that the transition probability naively appears to be unaffected by non-commutativity of space-time, but the mass squared difference has non-commutative corrections which enter the above equation through $\hat{\Delta}_{21}^{M}$ (see the discussion below eq.~\eqref{massdeformation2}). Hence, non-commutativity of space-time affects the neutrino oscillation probabilities of $2$-flavour neutrinos in the presence of matter and gravity. In the limit $a\rightarrow 0$, we get back the commutative result. 

\section{Neutrino Oscillations in $\kappa$-deformed space-time: Three flavour}

In this section, we derive neutrino oscillation probability for three flavours with the inclusion of gravitational effects, in constant matter density, in $\kappa$-deformed space-time. The effective Hamiltonian with constant matter density in mass basis is $\hat{H}_{M}=diag(\hat{E}_{1}, \hat{E}_{2}, \hat{E}_{3})+diag(V, 0,0)$, where $V=\sqrt{2}G_{F}N_{e}$ is the matter induced effective potential and $\hat{E}_{i}$ is given in eq.~\eqref{energyeigenvalues} but now, $i=1, 2,3$. After subtracting the common diagonal elements, as they do not affect the oscillation probabilities \cite{monojith} and subtracting $\frac{m_{1}^{2}}{2E}\left(1+\frac{ap^{0}}{2}\left(\frac{1}{1+\frac{m_{1}^{2}}{E^{2}}}\right)\right)+\frac{ap^{0}}{2}\frac{E}{1+\frac{m_{1}^{2}}{E^{2}}}$ from the diagonal elements gives,
\be \label{massbasisH}
\hat{H}_{M}^{(1)}=\frac{1}{2E}\begin{bmatrix}
0 & 0 & 0\\
0 & \Delta m_{21}^{2}\left(1-\frac{ap^{0}}{2\left(1+\frac{m_{1}^{2}}{E^{2}}\right)\left(1+\frac{m_{2}^{2}}{E^{2}}\right)}\right) & 0\\
0 & 0 & \Delta m_{31}^{2}\left(1-\frac{ap^{0}}{2\left(1+\frac{m_{1}^{2}}{E^{2}}\right)\left(1+\frac{m_{3}^{2}}{E^{2}}\right)}\right)
\end{bmatrix}+\begin{bmatrix}
V & 0 & 0\\
0 & 0 & 0\\
0 & 0 &0
\end{bmatrix}.
\ee
The interaction Hamiltonian for three flavours of neutrinos in $\kappa$-deformed space-time is, 
\be \label{interactionhamiltonian}
\hat{H}_{M}^{torsion}=\left(1+\frac{7}{3}ap^{0}+ap^{0}\left(\frac{2}{3}e_{\mu}^{~c}x_{l}\partial^{l}e_{c}^{~\mu}-\frac{2}{3}e_{\tilde{0}}^{~\mu}x_{l}\partial^{l}e_{\mu}^{~\tilde{0}}\right)\right)\tilde{n} ~diag(\lambda_{1}, \lambda_{2}, \lambda_{3}).
\ee 
On adding the above gravitational interaction Hamiltonian to eq.~\eqref{massbasisH}, we obtain,
\begin{multline} \label{massbasisH-gravity}
\hat{H}_{M}=\hat{H}_{M}^{(1)}+\left(1+\frac{7}{3}ap^{0}+ap^{0}\left(\frac{2}{3}e_{\mu}^{~c}x_{l}\partial^{l}e_{c}^{~\mu}-\frac{2}{3}e_{\tilde{0}}^{~\mu}x_{l}\partial^{l}e_{\mu}^{~\tilde{0}}\right)\right)\tilde{n}\begin{bmatrix}
0 & 0 & 0\\
0 & \Delta \lambda_{21} & 0\\
0 & 0 & \Delta \lambda_{31}
\end{bmatrix},
\end{multline}
where $\lambda_{1}$ has been subtracted from the diagonal elements of the torsion Hamiltonian. We write the above Hamiltonian as,
\be \label{massbasisH-gravity2}
\hat{H}_{M}=\frac{\Delta\hat{m}_{31}^{2}}{2E}\begin{bmatrix}
0 & 0 & 0\\
0 & \hat{\alpha} & 0\\
0 & 0 & 1
\end{bmatrix}+\begin{bmatrix}
V & 0 & 0\\
0 & 0 & 0\\
0 & 0 &0
\end{bmatrix},
\ee
Here $\hat{\alpha}=\frac{\Delta\hat{m}_{21}^{2}}{\Delta\hat{m}_{31}^{2}}$ and, 
\bea \label{masssquareddifference}
\Delta\hat{m}_{31}^{2}&=&\Delta \tilde{m}_{31}^{2}+ap^{0}\left(\frac{-\Delta m_{31}^{2}}{2\left(1+\frac{m_{1}^{2}}{E^{2}}\right)\left(1+\frac{m_{3}^{2}}{E^{2}}\right)}+\frac{2}{3}\tilde{n}E\Delta\lambda_{31}\left(7+2e_{\mu}^{~c}x_{l}\partial^{l}e_{c}^{~\mu}-2e_{\tilde{0}}^{~\mu}x_{l}\partial^{l}e_{\mu}^{~\tilde{0}}\right)\right), \\\nonumber
\Delta\hat{m}_{21}^{2}&=&\Delta \tilde{m}_{21}^{2}+ap^{0}\left(\frac{-\Delta m_{21}^{2}}{2\left(1+\frac{m_{1}^{2}}{E^{2}}\right)\left(1+\frac{m_{2}^{2}}{E^{2}}\right)}+\frac{2}{3}\tilde{n}E\Delta\lambda_{21}\left(7+2e_{\mu}^{~c}x_{l}\partial^{l}e_{c}^{~\mu}-2e_{\tilde{0}}^{~\mu}x_{l}\partial^{l}e_{\mu}^{~\tilde{0}}\right)\right),
\eea 
where $\Delta \tilde{m}_{31}^{2}=\Delta m_{31}^{2}+2\tilde{n}E\Delta\lambda_{31},~\Delta \tilde{m}_{21}^{2}=\Delta m_{21}^{2}+2\tilde{n}E\Delta\lambda_{21}$. Here, the first $ap^{0}$ dependent term is coming from the dispersion relation(eq.~\eqref{dispnrelatn2}) while the rest of the $ap^{0}$ is coming from the effect of non-commutativity in the gravitational interactions of fermions (see $\hat{H}_{M}^{torsion}$ in eq.~\eqref{interactionhamiltonian}). Using the leptonic mixing matrix \cite{akhmedov}, the Hamiltonian is transferred to the flavour basis with the inclusion of matter effects as,
\be \label{flavourbasisH}
\hat{H}_{F}=\frac{\Delta\hat{m}_{31}^{2}}{2E}U diag(0, \hat{\alpha}, 1)U^{\dagger}+diag (V, 0, 0).
\ee
This Hamiltonian can also be written as,
\be \label{flavourbasisH2}
\hat{H}_{F}=\frac{\Delta\hat{m}_{31}^{2}}{2E}O_{23}U_{\delta}\hat{M}U_{\delta}^{\dagger}O_{23}^{\dagger},
\ee
with $U_{\delta}=diag(1, 1, e^{-i\delta_{CP}})$, $\delta_{CP}$ is the Dirac type CP violating phase and $O_{ij}$ is the orthogonal rotation matrix in the $ij$ plane. In the above, $\hat{M}=O_{13}O_{12}diag(0, \hat{\alpha}, 1)O_{12}^{T}O_{13}^{T}+diag(\hat{A}, 0, 0)$ with the abbreviations $\hat{\Delta}=\frac{\Delta\hat{m}_{31}^{2}L}{4E}$ and $\hat{A}=\frac{2EV}{\Delta\hat{m}_{31}^{2}}=\frac{VL}{2\hat{\Delta}}$. The explicit form of the $\hat{M}$ matrix after keeping terms only upto second order in $\hat{\alpha}$ and $s_{13}$ is,
\be \label{Mmatrix} 
\hat{M}=\begin{bmatrix}
\hat{\alpha}s_{12}^{2}+s_{13}^{2}+\hat{A} & \hat{\alpha}s_{12}c_{12} & s_{13}-\hat{\alpha}s_{13}s_{12}^{2}\\
\hat{\alpha}c_{12}s_{12} & \hat{\alpha}c_{12}^{2} & -\hat{\alpha}s_{13}c_{12}s_{12}\\
s_{13}-\hat{\alpha}s_{13}s_{12}^{2} & -\hat{\alpha}s_{13}c_{12}s_{12} & 1-s_{13}^{2}
\end{bmatrix},
\ee
Here, $s_{ij}=sin\theta_{ij}$ and $c_{ij}=cos\theta_{ij}$; we have substituted $c_{13}^{2}=1-s_{13}^{2}$ and have assumed $c_{13}\approx 1$ as $\theta_{13}\leq 10.8^{\circ}$ \cite{akhmedov}. We can write $\hat{M}$ in terms of zeroth order, first order and second order matrices in $\hat{\alpha}$ and $s_{13}$ as $\hat{M}=\hat{M}^{(0)}+\hat{M}^{(1)}+\hat{M}^{(2)}$ where 
\be \label{M0matrix}
\hat{M}^{(0)}=\begin{bmatrix}
\hat{A} & 0 & 0\\
0 & 0 & 0\\
0 & 0 & 1
\end{bmatrix},~~\hat{M}^{(1)}=\begin{bmatrix}
\hat{\alpha}s_{12}^{2} & \hat{\alpha}s_{12}c_{12} & s_{13}\\
\hat{\alpha}c_{12}s_{12} & \hat{\alpha}c_{12}^{2} & 0\\
s_{13} & 0 & 0
\end{bmatrix},
\ee
\be \label{M2matrix}
\hat{M}^{(2)}=\begin{bmatrix}
s_{13}^{2} & 0 & -\hat{\alpha}s_{13}s_{12}^{2}\\
0 & 0 & -\hat{\alpha}s_{13}c_{12}s_{12}\\
-\hat{\alpha}s_{13}s_{12}^{2} & -\hat{\alpha}s_{13}c_{12}s_{12} & -s_{13}^{2}
\end{bmatrix}.
\ee
One finds the eigenvalues of $\hat{M}^{(0)}$, $\hat{M}^{(1)}$ and $\hat{M}^{(2)}$ using which the eigenvalues of the effective Hamiltonian is given as $\hat{E}_{i}=\frac{\Delta\hat{\tilde{m}}_{31}^{2}}{2E}\left(\hat{\Omega}_{i}^{(0)}+\hat{\Omega}_{i}^{(1)}+\hat{\Omega}_{i}^{(2)}\right)$ where,
\bea \label{Omega2}
\hat{\Omega}_{i}^{(1)}&=&\left<\hat{v}_{i}^{(0)}|\hat{M}^{(1)}|\hat{v}_{i}^{0}\right>, \\
\hat{\Omega}_{i}^{(2)}&=&\left<\hat{v}_{i}^{(0)}|\hat{M}^{(2)}|\hat{v}_{i}^{0}\right>+\sum_{j\neq i}\frac{\abs{\left<\hat{v}_{j}^{(0)}|\hat{M}^{(1)}|\hat{v}_{i}^{0}\right> }^{2}}{\left(\hat{\Omega}_{i}^{(0)}-\hat{\Omega}_{j}^{(0)}\right)},
\eea
and $\hat{\Omega}_{i}^{(0)}$ is the eigenvalue of $\hat{M}^{(0)}$. This gives the energy eigenvalues of the Hamiltonian as,
\bea \label{energyeigenvalues3} 
\hat{E}_{1}&=&\frac{\Delta\hat{m}_{31}^{2}}{2E}\left(\hat{A}+\hat{\alpha}s_{12}^{2}+\frac{\hat{A}s_{13}^{2}}{\hat{A}-1}+\frac{\hat{\alpha}^{2}sin^{2}2\theta_{12}}{4\hat{A}}\right),\\
\hat{E}_{2}&=&\frac{\Delta\hat{m}_{31}^{2}}{2E}\left(\hat{\alpha}c_{12}^{2}-\frac{\hat{\alpha}^{2}sin^{2}2\theta_{12}}{4\hat{A}}\right),\\
\hat{E}_{3}&=&\frac{\Delta\hat{m}_{31}^{2}}{2E}\left(1-\frac{\hat{A}s_{13}^{2}}{\hat{A}-1}\right).
\eea
Similarly, the eigenvectors of the $\hat{M}$ matrix can be written as $\hat{v}_{i}=\hat{v}_{i}^{(0)}+\hat{v}_{i}^{(1)}+\hat{v}_{i}^{(2)}$ where $\hat{v}_{i}^{(0)}$ is the eigenvector of $\hat{M}^{(0)}$ and,
\bea \label{eigenvectors1,2} 
\hat{v}_{i}^{(1)}&=&\sum_{j\neq i}\frac{\left<\hat{v}_{j}^{(0)}|\hat{M}^{(1)}|\hat{v}_{i}^{0}\right>}{\left(\hat{\Omega}_{i}^{(0)}-\hat{\Omega}_{j}^{(0)}\right)}\hat{v}_{j}^{(0)},\\
\hat{v}_{i}^{(2)}&=&\sum_{j\neq i}\frac{1}{\left(\hat{\Omega}_{i}^{(0)}-\hat{\Omega}_{j}^{(0)}\right)}\left[\left<\hat{v}_{j}^{(0)}|\hat{M}^{(2)}|\hat{v}_{i}^{0}\right>+\left(\hat{M}^{(1)}v_{i}^{(1)}\right)_{j}-\hat{\Omega}_{i}^{(1)}\left(\hat{v}_{i}^{(1)}\right)_{j}\right]\hat{v}_{j}^{(0)}.
\eea
Thus, the eigenvectors of $\hat{M}$ becomes,
\be \label{eigenvectorsM} 
\hat{v}_{1}=\begin{bmatrix}
1 \\ \frac{\hat{\alpha}sin2\theta_{12}}{2\hat{A}}+\frac{\hat{\alpha}^{2}sin4\theta_{12}}{4\hat{A}^{2}}\\ \frac{s_{13}}{\hat{A}-1}-\frac{\hat{\alpha}\hat{A}s_{13}s_{12}^{2}}{\left(\hat{A}-1\right)^{2}}
\end{bmatrix};~~~~\hat{v}_{2}=\begin{bmatrix}
 -\frac{\hat{\alpha}sin2\theta_{12}}{2\hat{A}}-\frac{\hat{\alpha}^{2}sin4\theta_{12}}{4\hat{A}^{2}}\\ 1 \\ \frac{\hat{\alpha}s_{13}sin2\theta_{12}(\hat{A}+1)}{2\hat{A}}
\end{bmatrix};~~~~\hat{v}_{3}=\begin{bmatrix}
-\frac{s_{13}}{\hat{A}-1}+\frac{\hat{\alpha}\hat{A}s_{13}s_{12}^{2}}{\left(\hat{A}-1\right)^{2}} \\ \frac{\hat{\alpha}\hat{A}s_{13}sin2\theta_{12}}{2(\hat{A}-1)}\\ 1
\end{bmatrix}.
\ee
The $\kappa$-deformed mixing matrix in matter becomes $\hat{U}^{\prime}=O_{23}U_{\delta}\hat{W}$ where $\hat{W}=(\hat{v}_{1}, \hat{v}_{2}, \hat{v}_{3})$. The general formula to find neutrino oscillation probabilities in deformed space-time is given as,
\begin{multline} \label{oscillationprobability}
\hat{P}_{\alpha\beta}=\delta_{\alpha\beta}-2\sum_{i<j}Re(\hat{U}^{\prime}_{\alpha i}\hat{U}^{\prime}_{\beta j}\hat{U}^{*\prime}_{\alpha j}\hat{U}^{*\prime }_{\beta i})\left[1-cos((\hat{E}_{i}-\hat{E}_{j})t)\right]+2\sum_{i<j}Im(\hat{U}^{\prime}_{\alpha i}\hat{U}^{\prime}_{\beta j}\hat{U}^{*\prime}_{\alpha j}\hat{U}^{*\prime }_{\beta i})sin((\hat{E}_{i}-\hat{E}_{j})t).
\end{multline}
This gives the following neutrino oscillation probabilities in $\kappa$-deformed space-time,
\be \label{Pee-3}
\hat{P}_{ee}=1-\frac{\hat{\alpha}^{2}sin^{2}2\theta_{12}}{\hat{A}^{2}}sin^{2}(\hat{\Delta}\hat{A})-\frac{4s_{13}^{2}}{(\hat{A}-1)^{2}}sin^{2}(\hat{\Delta}(\hat{A}-1)),
\ee
\bea \label{Pemu-3}
\hat{P}_{e\mu}&=&\frac{\hat{\alpha}^{2}c_{23}^{2}sin^{2}2\theta_{12}}{\hat{A}^{2}}sin^{2}(\hat{\Delta}\hat{A})+\frac{4s_{13}^{2}s_{23}^{2}}{(\hat{A}-1)^{2}}sin^{2}(\hat{\Delta}(\hat{A}-1))\\\nonumber
&+&\frac{2\hat{\alpha}s_{13}sin2\theta_{23}sin2\theta_{12}}{\hat{A}(\hat{A}-1)}cos(\hat{\Delta}-\delta_{CP})sin(\hat{\Delta}\hat{A})sin(\hat{\Delta}(\hat{A}-1)),
\eea
\bea \label{Ptau-3}
\hat{P}_{e\tau}&=&\frac{\hat{\alpha}^{2}s_{23}^{2}sin^{2}2\theta_{12}}{\hat{A}^{2}}sin^{2}(\hat{\Delta}\hat{A})+\frac{4s_{13}^{2}c_{23}^{2}}{(\hat{A}-1)^{2}}sin^{2}(\hat{\Delta}(\hat{A}-1))\\\nonumber
&-&\frac{2\hat{\alpha}s_{13}sin2\theta_{23}sin2\theta_{12}}{\hat{A}(\hat{A}-1)}cos(\hat{\Delta}-\delta_{CP})sin(\hat{\Delta}\hat{A})sin(\hat{\Delta}(\hat{A}-1)),
\eea
\bea \label{Pmumu-3}
\hat{P}_{\mu\mu}&=&1-sin^{2}2\theta_{23}sin^{2}\hat{\Delta}+\hat{\alpha}c_{12}^{2}sin^{2}2\theta_{23}\hat{\Delta}sin(2\hat{\Delta})-\frac{\hat{\alpha}^{2}c_{23}^{2}sin^{2}2\theta_{12}}{\hat{A}^{2}}sin^{2}(\hat{\Delta}\hat{A})\\\nonumber
&-&\hat{\alpha}^{2}c_{12}^{4}sin^{2}2\theta_{23}\hat{\Delta}^{2}cos2\hat{\Delta}
-\frac{4s_{13}^{2}s_{23}^{2}}{(\hat{A}-1)^{2}}sin^{2}(\hat{\Delta}(\hat{A}-1))\\\nonumber
&+&\frac{1}{2\hat{A}}\hat{\alpha}^{2}sin^{2}2\theta_{23}sin^{2}2\theta_{12}\left(sin(\hat{\Delta}\hat{A})\frac{sin\hat{\Delta}}{\hat{A}}cos(\hat{\Delta}(\hat{A}-1))-\frac{\hat{\Delta}}{2}sin2\hat{\Delta}\right)\\\nonumber
&-&\frac{2}{(\hat{A}-1)}s_{13}^{2}sin^{2}2\theta_{23}\left(cos(\hat{\Delta}\hat{A})\frac{sin\hat{\Delta}}{(\hat{A}-1)}sin(\hat{\Delta}(\hat{A}-1))-\frac{\hat{\Delta}}{2}\hat{A}sin2\hat{\Delta}\right)\\\nonumber
&-&\frac{2\hat{\alpha}s_{13}sin2\theta_{12}sin2\theta_{23}}{\hat{A}(\hat{A}-1)}cos\delta_{CP}sin(\hat{\Delta}\hat{A})sin(\hat{\Delta}(\hat{A}-1))cos\hat{\Delta}\\\nonumber
&+&\frac{2}{(\hat{A}-1)}\hat{\alpha}s_{13}sin2\theta_{12}sin2\theta_{23}cos2\theta_{23}cos\delta_{CP}\left(\hat{A}sin^{2}\hat{\Delta}-\frac{sin\hat{\Delta}sin(\hat{\Delta}\hat{A})}{\hat{A}}cos(\hat{\Delta}(\hat{A}-1))\right),
\eea
and 
\bea \label{Pmutau-3}
\hat{P}_{\mu\tau}&=&sin^{2}2\theta_{23}sin^{2}\hat{\Delta}-\hat{\alpha}c_{12}^{2}sin^{2}2\theta_{23}\hat{\Delta}sin2\hat{\Delta}+\hat{\alpha}^{2}c_{12}^{4}sin^{2}2\theta_{23}\hat{\Delta}^{2}cos2\hat{\Delta}\\\nonumber
&-&\frac{\hat{\alpha}^{2}}{2\hat{A}}sin^{2}2\theta_{23}sin^{2}2\theta_{12}\left(sin\hat{\Delta}\frac{sin(\hat{\Delta}\hat{A})}{\hat{A}}cos(\hat{\Delta}(\hat{A}-1))-\frac{\hat{\Delta}}{2}sin2\hat{\Delta}\right)\\\nonumber
&+&\frac{2}{(\hat{A}-1)}s_{13}^{2}sin^{2}2\theta_{23}\left(sin\hat{\Delta}cos(\hat{\Delta}\hat{A})\frac{sin(\hat{\Delta}(\hat{A}-1))}{(\hat{A}-1)}-\frac{\hat{\Delta}\hat{A}}{2}sin2\hat{\Delta}\right)\\\nonumber
&+&\frac{2\hat{\alpha}s_{13}sin2\theta_{23}sin2\theta_{12}}{\hat{A}(\hat{A}-1)}sin\delta_{CP}sin\hat{\Delta}sin(\hat{\Delta}\hat{A})sin(\hat{\Delta}(\hat{A}-1))\\\nonumber
&-&\frac{2\hat{\alpha}s_{13}sin2\theta_{23}sin2\theta_{12}}{(\hat{A}-1)}cos\delta_{CP}cos2\theta_{23}sin\hat{\Delta}\left(\hat{A}sin\hat{\Delta}-\frac{sin(\hat{\Delta}\hat{A})}{\hat{A}}cos(\hat{\Delta}(\hat{A}-1))\right).
\eea

Note that in the limit $a\rightarrow 0$, we get back the commutative neutrino oscillation probabilities for $3$ flavour neutrinos \cite{amitabha}.

\section{Neutrino Oscillation probabilities of terrestrial neutrinos in $\kappa$-deformed Schwarzschild background}

In this section, we look at oscillations of terrestrial muon neutrinos in $\kappa$-deformed Schwarzschild space-time. The Schwarzschild metric given as,
\be \label{schwarschild}
ds^{2}=-\left(1-\frac{2GM}{rc^{2}}\right)c^{2}dt^{2}+\left(1-\frac{2GM}{rc^{2}}\right)^{-1}dr^{2}+r^{2}\left(d\theta^{2}+sin^{2}\theta d\phi^{2}\right).
\ee 
The corresponding tetrads and inverse tetrads of the above metric are,
\bea \label{tetrads-schwarschild}
e_{0}^{~\tilde{0}}=\left(1-\frac{2GM}{rc^{2}}\right)^{\frac{1}{2}};~~&&~~e_{1}^{~\tilde{1}}=\left(1-\frac{2GM}{rc^{2}}\right)^{-\frac{1}{2}}\\
e_{2}^{~\tilde{2}}=r;~~&&~~e_{3}^{~\tilde{3}}=rsin\theta.
\eea
\bea \label{inversetetrads-schwarschild} 
e_{\tilde{0}}^{~0}=\left(1-\frac{2GM}{rc^{2}}\right)^{-\frac{1}{2}};~~&&~~e_{\tilde{1}}^{~1}=\left(1-\frac{2GM}{rc^{2}}\right)^{\frac{1}{2}}\\
e_{\tilde{2}}^{~2}=\frac{1}{r};~~&&~~e_{\tilde{3}}^{~3}=\frac{1}{rsin\theta}.
\eea
Thus, mass squared difference (eq.~\eqref{masssquareddifference}) reduces to,
\bea \label{masssquareddifference-schwarschild}
\Delta\hat{m}_{31}^{2}&=&\Delta \tilde{m}_{31}^{2}+ap^{0}\left(\frac{-\Delta m_{31}^{2}}{2\left(1+\frac{m_{1}^{2}}{E^{2}}\right)\left(1+\frac{m_{3}^{2}}{E^{2}}\right)}+\frac{2}{3}\tilde{n}E\Delta\lambda_{31}\left(3-\frac{2GM}{rc^{2}}\left[1-\frac{2GM}{rc^{2}}\right]^{-1}\right)\right), \\\nonumber
\Delta\hat{m}_{21}^{2}&=&\Delta \tilde{m}_{21}^{2}+ap^{0}\left(\frac{-\Delta m_{21}^{2}}{2\left(1+\frac{m_{1}^{2}}{E^{2}}\right)\left(1+\frac{m_{2}^{2}}{E^{2}}\right)}+\frac{2}{3}\tilde{n}E\Delta\lambda_{21}\left(3-\frac{2GM}{rc^{2}}\left[1-\frac{2GM}{rc^{2}}\right]^{-1}\right)\right).
\eea 

In order to consider oscillations of terrestrial muon neutrinos in Earth's crust, the average matter density of Earth's crust is taken as $2.7g/cm^{3}$ and the neutrino energy range is taken as $0.5GeV < E < 10GeV$. This is appropriate for DUNE experiment with baseline $L=1300km$. Other chosen values are $\Delta m_{21}^{2}= 7.39 \times 10^{-23}GeV^{2}$, $\Delta m_{31}^{2}= 2.454 \times 10^{-21}GeV^{2}$, $\theta_{12}=33.8^{\circ}$, $\theta_{23}=48.6^{\circ}$, $\theta_{13}=8.6^{\circ}$ and $\delta_{CP}=221^{\circ}$ \cite{workman}. For the terms coming from the Schwarzschild metric in eq.~\eqref{masssquareddifference-schwarschild}, we have chosen, $G=6.67 \times 10^{-11}Nm^{2}kg^{-2}$, $M=6 \times 10^{24}kg$ as the mass of earth and $r=6.371 \times 10^{6}m$ as the radius of the earth. Also, $\tilde{n}=\lambda n$ and $\lambda\Delta\lambda_{31}=\lambda\Delta\lambda_{21}=\Delta\lambda^{2}$ which is taken as $0.1G_{F}$ and $1G_{F}$ respectively\cite{amitabha}.

\begin{figure}
\centering
    \includegraphics[width=0.75\textwidth]{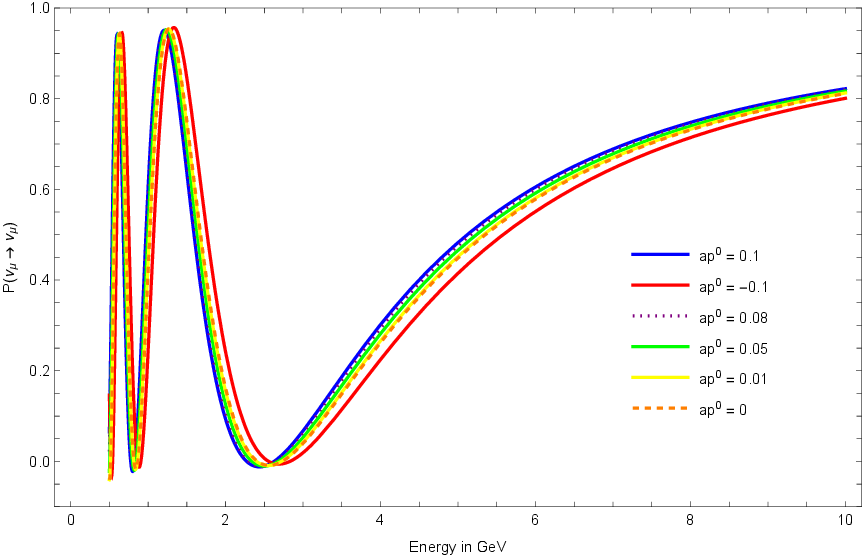}
    \caption{$\hat{P}_{\mu\mu}$ v/s $E$ is plotted for $\Delta\lambda^{2}=1G_{F}$ for different values of $ap^{0}$.} 
    \label{pmumu-ap0variation}
\end{figure}

\begin{figure}
\centering
    \includegraphics[width=0.75\textwidth]{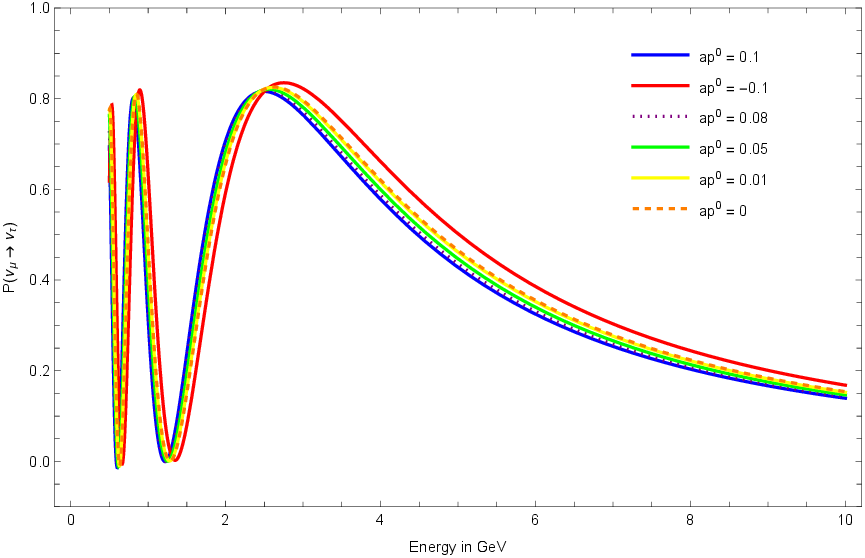}
    \caption{$\hat{P}_{\mu\tau}$ v/s $E$ is plotted for $\Delta\lambda^{2}=1G_{F}$ for different values of $ap^{0}$.} 
    \label{pmutau-ap0variation}
\end{figure}

Using these, we plot neutrino oscillation probabilities of muon neutrinos in DUNE experiment. The $\nu_{\mu} \rightarrow \nu_{\mu}$ survival probability in $\kappa$-deformed space-time is plotted against $E$ ranging from $0.5GeV$ to $10GeV$. Since mass of neutrinos is in the eV scale and its energy is in the GeV scale, the denominator of the second term i.e. $\left(1+\frac{m_{1}^{2}}{E^{2}}\right)\left(1+\frac{m_{2}^{2}}{E^{2}}\right)$ in eq.~\eqref{masssquareddifference-schwarschild} is taken as $1$. In Fig.(\ref{pmumu-ap0variation}), $\Delta\lambda^{2}$ is fixed as $1G_{F}$ and $\hat{P}_{\mu\mu}$ is plotted against $E$ for various values of $ap^{0}$.
We see that as the non-commutativity of space-time increases, the transition probability of $\nu_{\mu}\rightarrow \nu_{\mu}$ also increases for higher values of energy. The oscillation probability is seen to increase from the commutative case for positive values of $ap^{0}$ and decrease for negative values of $ap^{0}$. The highest deviation is seen for $ap^{0}$ being 0.1. Next, we plot $\hat{P}_{\mu\tau}$ against $E$ for $\Delta\lambda^{2}=1G_{F}$ for different values of $ap^{0}$ in Fig.(\ref{pmutau-ap0variation}).
The behaviour seen in Fig.(\ref{pmutau-ap0variation}) is opposite to that of Fig.(\ref{pmumu-ap0variation}) and the oscillation probability increases from the commutative case for $ap^{0}$ being negative and decreases from the commutative case for $ap^{0}$ being positive. Finally, transition probability of $\mu \rightarrow e$ is plotted against $E$ for $\Delta\lambda^{2}=1G_{F}$ for different values of $ap^{0}$. The behaviour of the plot is similar to the one seen in Fig.(\ref{pmumu-ap0variation}).
\begin{figure}
\centering
    \includegraphics[width=0.75\textwidth]{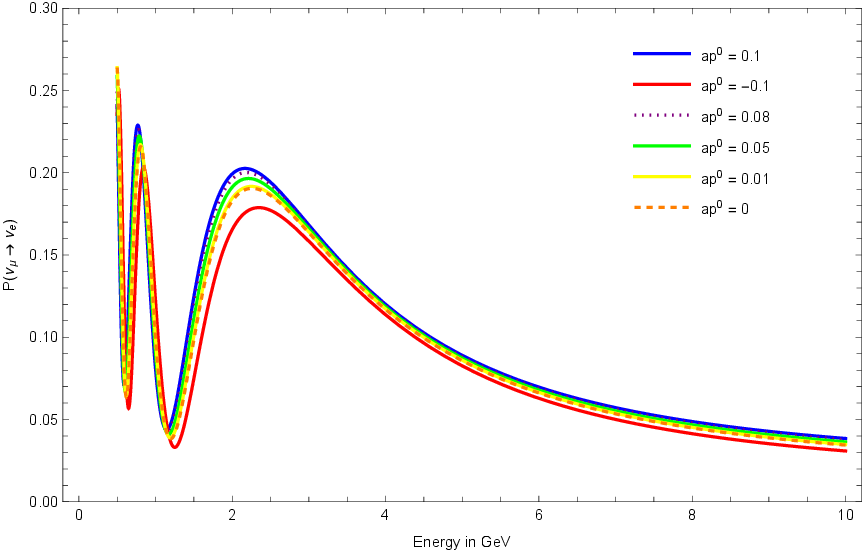}
    \caption{$\hat{P}_{\mu e}$ v/s $E$ is plotted for $\Delta\lambda^{2}=1G_{F}$ for different values of $ap^{0}$.} 
    \label{pmue-ap0variation}
\end{figure} 

\begin{figure}
\centering
    \includegraphics[width=0.75\textwidth]{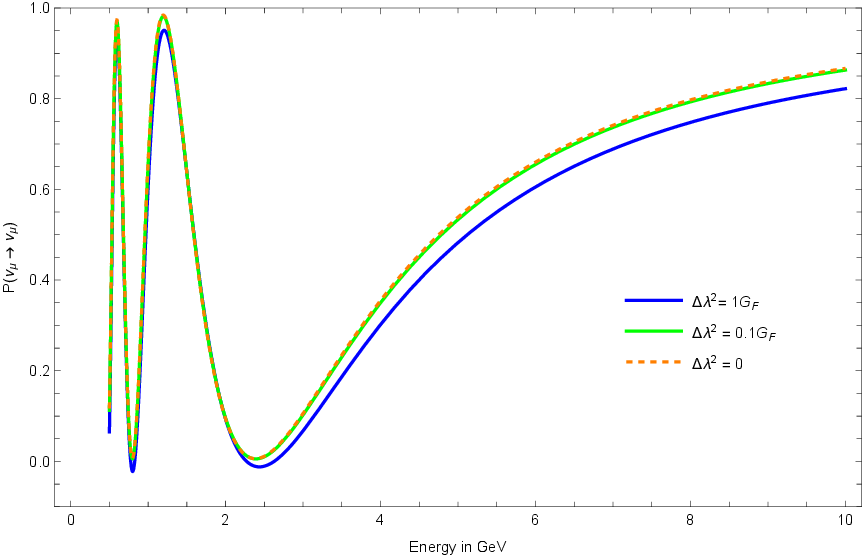}
    \caption{$\hat{P}_{\mu\mu}$ v/s $E$ is plotted for $ap^{0}=0.1$ for different values of $\Delta\lambda^{2}$.} 
    \label{pmumu-lambdavariation}
\end{figure}

\begin{figure}
\centering
    \includegraphics[width=0.75\textwidth]{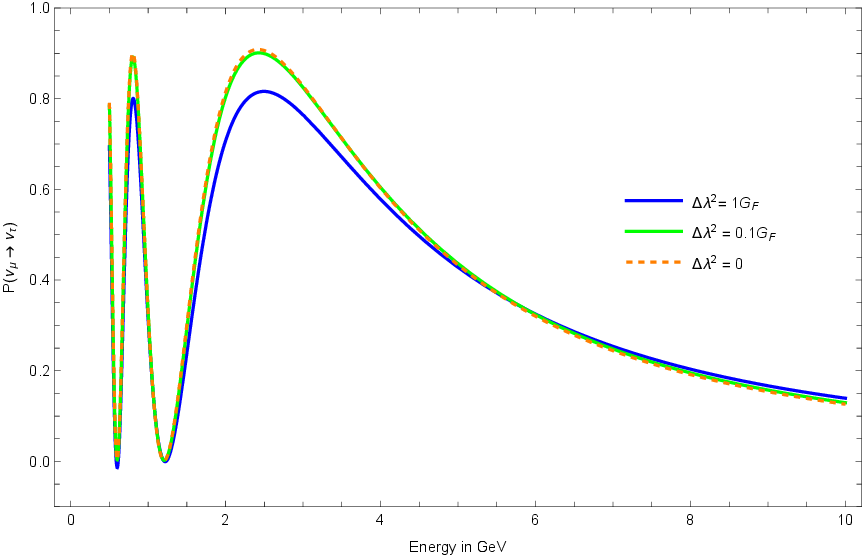}
    \caption{$\hat{P}_{\mu\tau}$ v/s $E$ is plotted for $ap^{0}=0.1$ for different values of $\Delta\lambda^{2}$.} 
    \label{pmutau-lambdavariation}
\end{figure}

\begin{figure}
\centering
    \includegraphics[width=0.75\textwidth]{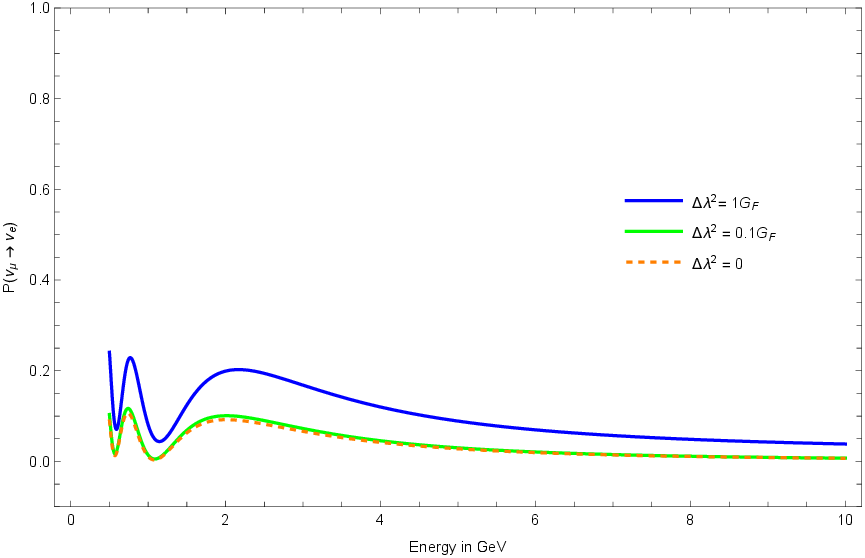}
    \caption{$\hat{P}_{\mu e}$ v/s $E$ is plotted for $ap^{0}=0.1$ for different values of $\Delta\lambda^{2}$.} 
    \label{pmue-lambdavariation}
\end{figure}

Next, we fix $ap^{0}$ as $0.1$ and the survival probability for $\nu_{\mu}\rightarrow \nu_{\mu}$ is plotted against $E$ by varying $\Delta\lambda^{2}$ in Fig.(\ref{pmumu-lambdavariation}). 
We see the most deviation for $\Delta\lambda^{2}$ being $1G_{F}$ and the survival probability is found to decrease as $\Delta\lambda^{2}$ increases for higher values of energy. This behaviour is similar to the one seen in the commutative space-time\cite{amitabha}. The variation of transition probability of $\nu_{\mu}\rightarrow \nu_{\tau}$ against $E$ for $ap^{0}=0.1$ and varying $\Delta\lambda^{2}$ is plotted in Fig.(\ref{pmutau-lambdavariation}). The behaviour of this plot is similar to that of $\hat{P}_{\mu\mu}$ in Fig.(\ref{pmumu-lambdavariation}).
Next, $\hat{P}_{\mu e}$ is plotted against $E$ for $ap^{0}=0.1$ and by varying $\Delta\lambda^{2}$ in Fig.(\ref{pmue-lambdavariation}). Here, we see that the deviation in the transition probability increases as the value of torsional coupling constant $\Delta\lambda^{2}$ is increased. This behaviour is opposite to the similar plots for $\hat{P}_{\mu\mu}$ and $\hat{P}_{\mu e}$ (see Fig.(\ref{pmumu-lambdavariation}), Fig.(\ref{pmutau-lambdavariation})). 

\begin{figure}
\centering
    \includegraphics[width=0.75\textwidth]{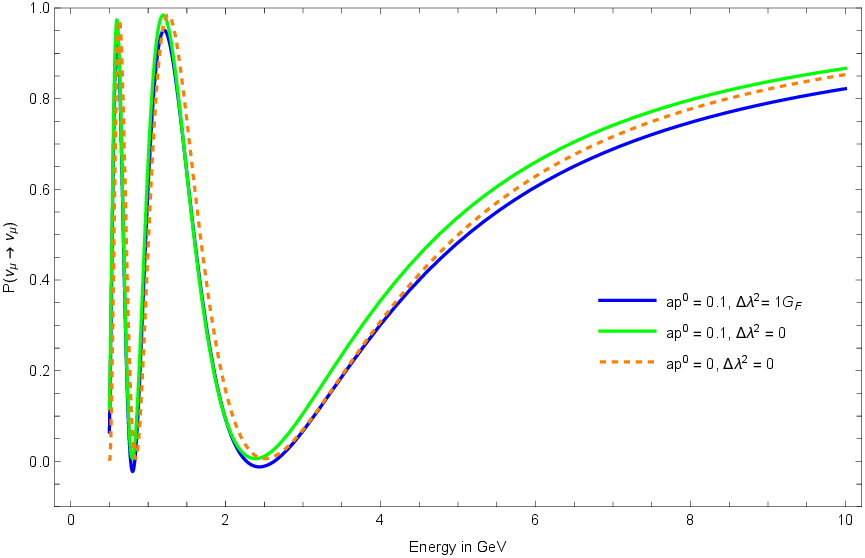}
    \caption{$\hat{P}_{\mu\mu}$ v/s $E$ is plotted for $3$ cases.} 
    \label{pmumu-3}
\end{figure}

\begin{figure}
\centering
    \includegraphics[width=0.75\textwidth]{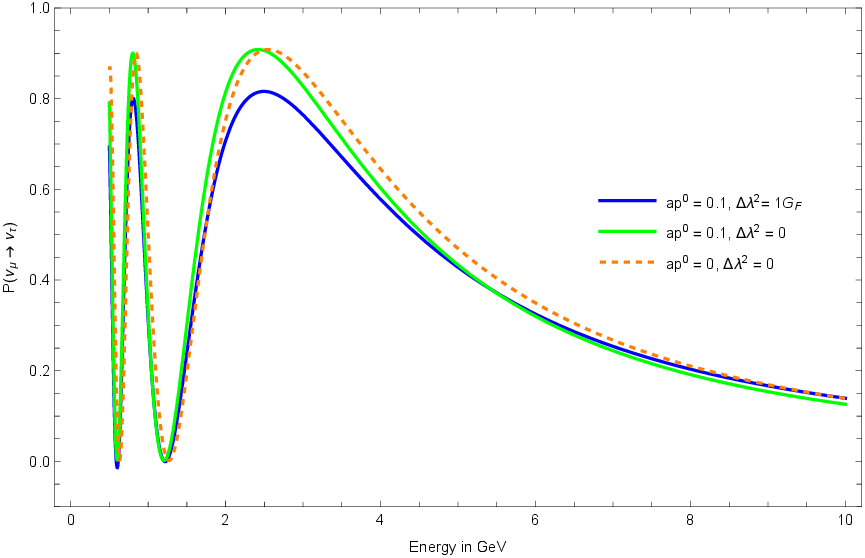}
    \caption{$\hat{P}_{\mu\tau}$ v/s $E$ is plotted for $3$ cases.} 
    \label{pmutau-3}
\end{figure}

\begin{figure}
\centering
    \includegraphics[width=0.75\textwidth]{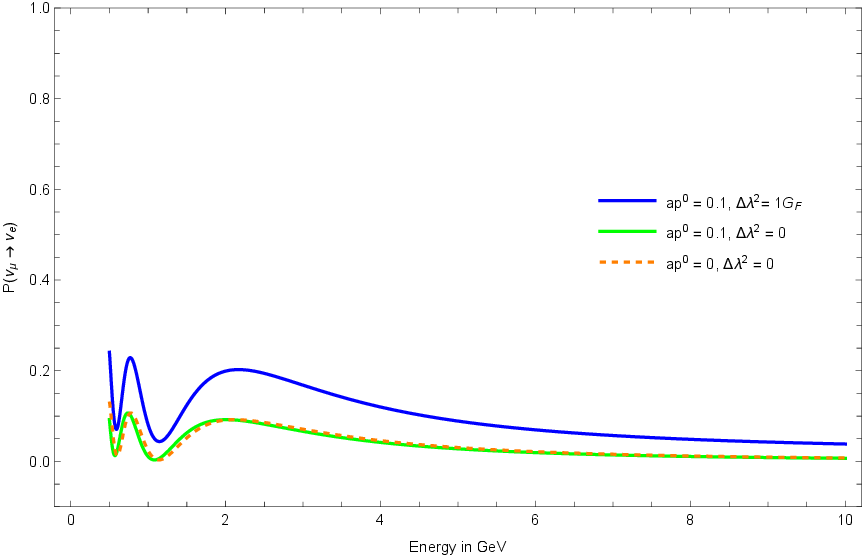}
    \caption{$\hat{P}_{\mu e}$ v/s $E$ is plotted for $3$ cases.} 
    \label{pmue-3}
\end{figure}

Fig.(\ref{pmumu-3}) is plotted for $\hat{P}_{\mu\mu}$ against E for three different cases, $(i)$ for $ap^{0}=0.1$ in presence of torsion, $(ii)$ for $ap^{0}=0.1$ in the absence of torsion and $(iii)$ in the absence of torsion and non-commutativity of space-time. Note that in the absence of torsional effects eq.~\eqref{masssquareddifference-schwarschild} still has non-commutative correction appearing from the deformed dispersion relation.
In Fig.(\ref{pmumu-3}), we see that the transition probability increases in the absence of torsion compared to the one in the absence of torsion and non-commutativity of space-time, while the oscillation probability decreases in the presence of torsion and non-commutativity of space-time compared to the one in the absence of torsion and non-commutativity of space-time. Similar graph is plotted for $\hat{P}_{\mu\tau}$ against $E$ in Fig.(\ref{pmutau-3}).
Fig.(\ref{pmue-3}) shows variation of $\hat{P}_{\mu e}$ against $E$ $(i)$ for $ap^{0}=0$ in presence of torsion, $(ii)$ for $ap^{0}=0.1$ in the absence of torsion and, $(iii)$ in the absence of torsion and non-commutativity of space-time.

From all these plots, we see that the effect of non-commutativity of space-time and torsion is prominent for $ap^{0}=0.1$ and $\Delta\lambda^{2}=1G_{F}$ for larger values of energy.

\section{Conclusion}

We have studied the effect of non-commutativity of space-time on neutrino oscillation probabilities when gravitational interactions of propagating neutrinos are incorporated. In order to do so, we have first looked at the effect of $\kappa$-deformed space-time on propagating fermions under the influence of gravitation. Gravity is incorporated into the action of fermions using the spin connection which involves the universal torsion free Levi-Civita connection and the contortion tensor to incorporate the effects of torsion. This action, which involves gravity and the fermionic Lagrangian, is all generalised to $\kappa$-deformed space-time. Using Taylor series expansion and equations for deformed tetrads, gamma matrices and Dirac derivative, all the non-commutative quantities in the action are written in terms of commutative quantities and their deformations. Further, the action is varied in order to find the equation satisfied by the contortion tensor which is put back into the action and further simplified. Then the corresponding Dirac equation is derived in $\kappa$-deformed space-time. These steps are repeated and generalised to incorporate all the species of fermions. When all species of fermions are considered, the contortion tensor is considered to couple to different species with different coupling strengths, hence the torsional coupling constants are introduced in the equations. A four-fermion interaction term with deformations due to non-commutativity of space-time is seen to arise in the effective Lagrangian due to gravitational interactions of the propagating fermions. The deformations due to non-commutativity of space-time on the interaction terms are seen to depend on the background metric through tetrads. 

Now, we use this interaction term to see how they affect oscillation probabilities for $2$ and $3$ flavour of neutrinos. Two-flavour neutrino oscillations are studied in both vacuum and in the presence of matter and gravity in $\kappa$-deformed space-time. The two-flavour neutrino oscillations in vacuum are found to be modified because of the modification in the dispersion relation in $\kappa$-deformed space-time. The gravitational interaction term from the Lagrangian is incorporated into the Hamiltonian along with the matter effects while studying two-flavour neutrino oscillations in the presence of matter and gravity in deformed space-time. We see that the mass squared difference is modified due to torsional effects and quantisation of space-time which ultimately lead to modifications to the two-flavour neutrino oscillation probabilities.

Similarly, the four-fermion interaction term is incorporated to find the effective Hamiltonian which is used to determine the three-flavour neutrino oscillation probabilities. We see that the mass squared difference is modified due to torsional effects and non-commutativity of space-time which in turn affects the neutrino oscillation probabilities. The modifications on the mass squared difference due to non-commutativity of space-time are found to depend on the background metric via the tetrads. All the obtained results in this paper revert back to the commutative result in \cite{amitabha} in the limit $a \rightarrow 0$.

We have plotted the obtained neutrino oscillation probabilities for terrestrial neutrinos in the background Schwarzschild metric, considering experimental measurements appropriate to the DUNE experiment. We see the maximum modification from the commutative result for $ap^{0}=0.1$ and $\Delta\lambda^{2}=1G_{F}$. We have plotted the transition probabilities for $\nu_{\mu}\rightarrow\nu_{e}$, $\nu_{\mu}\rightarrow\nu_{\tau}$ and survival probability for $\nu_{\mu}\rightarrow\nu_{\mu}$ against energy where $E$ ranges from $0.5GeV$ to $10GeV$. Three kinds of plots are presented for all the aforementioned oscillation probabilities. The first kind of graph is plotted by fixing the torsional coupling constant to be $\Delta\lambda^{2}=1G_{F}$ and varying the deformation parameter from $-0.1$ to $0.1$. The least deviation was seen for $ap^{0}=0.01$ and the behaviour of $\hat{P}_{\mu\tau}$(see Fig.(\ref{pmutau-ap0variation})) is opposite to that of $\hat{P}_{\mu\mu}$ (see Fig.(\ref{pmumu-ap0variation})) and $\hat{P}_{\mu e}$ (see Fig.(\ref{pmue-ap0variation})) as energy increases. Second type of graphs plotted was by fixing $ap^{0}=0.1$ and varying the torsional coupling constant for all the oscillation probabilities. We found the most deviation for $\Delta\lambda^{2}=1G_{F}$ in all three cases which is similar to the result obtained in \cite{amitabha}. Thirdly, we plotted all three oscillation probabilities against energy for $ap^{0}=0.1$ with and without torsion and in the absence of torsion and non-commutativity of space-time.

We see that quantum gravity effects are prominent when the effect of gravity is incorporated into neutrinos propagating with high energies and could be relevant in future measurements of neutrino oscillation probability experiments. Study of quantum decoherence for $B$ meson system has been done in \cite{dhiren} and they look at how it affects the fundamental parameters. Neutrino decoherence in the framework of non-commutative space-time has been studied recently \cite{nandi} and it will be interesting to see how curvature effects modify this results.

\section{Acknowledgement}
We thank Prof.Rukmani Mohanta for useful discussions and suggestions and Dr.Monojit Ghosh, Dr.Papia Panda and Ms.Priya Mishra for useful discussions. HS thanks Prime Minister Research Fellowship (PMRF id:3703690) for the financial support.

\renewcommand{\thesection}{Appendix-A:}
\section{Deformation of tetrads and gamma matrices.}
\renewcommand{\thesection}{A}
A metric in a curved space-time can be written in terms of an orthonormal set of basis vectors ($\hat{e}^{a}$) in the corresponding tangent space of the Minkowski space-time at a point $\hat{x}_{\mu}$. The generic form of a $\kappa$-deformed metric \cite{zuhair} can be written in terms of the orthonormal basis vectors as,
\bea \label{deformedmetric}
d\hat{s}^2&=&g_{00} dx^{0}dx^{0}+g_{ij}e^{-4ap^{0}}dx^{i}dx^{j}\\ \nonumber
&=&\hat{\eta}_{ab} ~\hat{e}^{a} \otimes \hat{e}^{b},
\eea
where $\hat{\eta}_{ab}=(-1, e^{-2ap^{0}}, e^{-2ap^{0}}, e^{-2ap^{0}})$. The Greek indices $\mu, \nu$ are for quantities in the curved space-time and indices $a, b$ are for those in the flat space-time. Here, $dx^{0}, dx^{i}$ on the RHS of eq.~\eqref{deformedmetric} are infinitesimal increments of commutative space-time coordinates. Thus,
\bea
d\hat{s}^{2}&=&g_{00}dx^{0}dx^{0}+g_{11}dx^{1}dx^{1}+g_{22}dx^{2}dx^{2}+g_{33}dx^{3}dx^{3}\\ \nonumber
&=&-\hat{e}^{0} \hat{e}^{0}+\hat{e}^{1} \hat{e}^{1} e^{-2ap^{0}}+\hat{e}^{2} \hat{e}^{2} e^{-2ap^{0}}+\hat{e}^{3} \hat{e}^{3} e^{-2ap^{0}}.
\eea
From the above relation, the deformed basis vectors in terms of the components of the metric tensor are obtained as,
\bea \label{deformedbasisvectors}
\hat{e}^{\tilde{0}}=\sqrt{g_{00}}dx^{0},&&~~~~\hat{e}^{\tilde{1}}=\sqrt{g_{11}}e^{-ap^{0}}dx^{1},\\ \nonumber
\hat{e}^{\tilde{2}}=\sqrt{g_{22}}e^{-ap^{0}}dx^{2},&&~~~~\hat{e}^{\tilde{3}}=\sqrt{g_{33}}e^{-ap^{0}}dx^{3}.
\eea
Here, indices $\tilde{0}, \tilde{1}$ etc. refer to components of coordinates of the flat deformed space-time and $0, 1$ etc. refers those of the curved deformed space-time.
We have $\hat{e}^{a}=\hat{e}_{0}^{~a}dx^{0}$ and $\hat{e}^{a}=\hat{e}_{i}^{~a}e^{-ap^{0}}dx^{i}$ which gives the deformed tetrads as,
\bea \label{deformedtetrads}
\hat{e}_{0}^{~\tilde{0}}=\sqrt{g_{00}},&&~~~~\hat{e}_{1}^{~\tilde{1}}=\sqrt{g_{11}},\\ \nonumber
\hat{e}_{2}^{~\tilde{2}}=\sqrt{g_{22}},&&~~~~\hat{e}_{3}^{~\tilde{3}}=\sqrt{g_{33}}.
\eea
The inverse of the deformed tetrads is found from the relation $\hat{e}_{a}^{~\mu}=\hat{g}^{\mu\nu}\hat{\eta}_{ab}\hat{e}_{\nu}^{~b}$ as,
\bea \label{inversetetrads}
\hat{e}_{\tilde{0}}^{~0}=e_{\tilde{0}}^{~0}=g^{00}\sqrt{g_{00}},&&~~~~\hat{e}_{\tilde{1}}^{~1}=e_{\tilde{1}}^{~1}e^{2ap^{0}}=g^{11}\sqrt{g_{11}}e^{2ap^{0}},\\ \nonumber
\hat{e}_{\tilde{2}}^{~2}=e_{\tilde{2}}^{~2}e^{2ap^{0}}=g^{22}\sqrt{g_{22}}e^{2ap^{0}},&&~~~~\hat{e}_{\tilde{3}}^{~3}=e_{\tilde{3}}^{~3}e^{2ap^{0}}=g^{33}\sqrt{g_{33}}e^{2ap^{0}}.
\eea
Note that the non-commutative tetrads are all related to the corresponding commutative ones by $ap^{0}$ dependent multiplicative factors except $\hat{e}_{\tilde{0}}^{~0}$ which is not modified.

In order to find the deformed gamma matrices, $\hat{\gamma}^{\mu}$, we consider the Clifford algebra in $\kappa$-deformed space-time as $[\hat{\gamma}^{a}, \hat{\gamma}^{b}]=2\hat{\eta}^{ab}$. Since $\hat{\eta}_{00}$ has no deformation, we find, $\hat{\gamma}^{0}= \gamma^{0}$ and using the space components of the $\kappa$-Minkowski metric and the Clifford algebra, we find, $\hat{\gamma}^{i}=e^{ap^{0}}\gamma^{i}$. Similarly, the inverse of the gamma matrices are obtained as $\hat{\gamma}_{0}=\gamma_{0}$ and $\hat{\gamma_{i}}=e^{-ap^{0}}\gamma^{i}$. The deformed gamma matrices in curved space-time are written using tetrads as $\hat{\gamma}^{\mu}=\hat{e}_{a}^{~\mu}\hat{\gamma}^{a}$.

\end{document}